\documentclass[12pt,preprint]{aastex}

\shorttitle{Preliminary Orbit for Haro 1-14c}
\shortauthors{Schaefer et al.}

\begin{document}

\title{Preliminary Orbit of the Young Binary Haro 1-14c}

\author{G. H. Schaefer\altaffilmark{1,2}, M. Simon\altaffilmark{3}, L. Prato\altaffilmark{4}, and T. Barman\altaffilmark{4}}

\altaffiltext{1}{The CHARA Array of Georgia State University, Mount Wilson Observatory, Mount Wilson, CA 91023; schaefer@chara-array.org}
\altaffiltext{2}{Space Telescope Science Institute, 3700 San Martin Drive, Baltimore, MD 21218}
\altaffiltext{3}{Department of Physics \& Astronomy, SUNY Stony Brook, Stony Brook, NY 11794-3800}
\altaffiltext{4}{Lowell Observatory, 1400 West Mars Hill Road, Flagstaff, AZ 86001}

\begin{abstract}

Using the Keck Interferometer, we spatially resolved the orbit of the pre$-$main-sequence binary, Haro 1-14c, for the first time.  We present these interferometric observations along with additional spectroscopic radial velocity measurements of the components.  We performed a simultaneous orbit fit to the interferometric visibilities and the radial velocities of Haro 1-14c.  Based on a statistical analysis of the possible orbital solutions that fit the data, we determined component masses of $M_1 = 0.96^{+0.27}_{-0.08}~M_{\odot}$ and $M_2 = 0.33^{+0.09}_{-0.02}~M_{\odot}$ for the primary and secondary, respectively, and a distance to the system of $111^{+19}_{-18}$~pc.  The distance measurement is consistent with the close distance estimates of the Ophiuchus molecular cloud.  Comparing our results with evolutionary tracks suggests an age of 3-4 Myr for Haro 1-14c.  With additional interferometric measurements to improve the uncertainties in the masses and distance, we expect the low-mass secondary to provide important empirical data for calibrating the theoretical evolutionary tracks for pre$-$main-sequence stars.

\end{abstract}


\section{Introduction}

For masses smaller than $\sim1 M_{\odot}$, uncertainties in the pre$-$main-sequence (PMS) evolutionary tracks are sufficiently great that the usual method of estimating the mass and age of a young star, by its location in the H-R diagram, can produce a scatter as large as a factor of 2 in mass and 3 in age depending on the tracks used \citep[e.g.,][]{simon01,mathieu07}.  As a result, the mass spectrum of stars produced in a star-forming region, the distribution of masses in binaries, and the region's star-forming history are imprecisely known.  Measuring reliable masses, luminosities, and effective temperatures of young stars contributes empirical data for testing calculations of young star evolution.  Currently, there are few PMS stars with reliably measured masses; in particular, there are none below 0.4~$M_\odot$ with a precision better than 10\% \citep{hillenbrand04, mathieu07}.

Mapping the orbital motion of a binary system provides a way to measure the dynamical masses of PMS stars.  Observations of a system as both a visual and spectroscopic binary yield the component stellar masses as well as the distance to the system.  Spectroscopic radial velocity measurements of PMS binaries are typically limited to systems which have periods smaller than a few years.  At the distance of nearby star forming regions like Taurus and Ophiuchus, this range of orbital periods corresponds to angular separations smaller than a few tens of milli-arcseconds (mas).  As a result, there are few PMS spectroscopic binaries that can be resolved spatially using the current generation of high resolution instrumentation.  \citet{steffen01} applied this approach to the PMS binary NTTS 045251+3016 by combining single- and double-lined radial velocity measurements with spatially resolved observations obtained with the Fine Guidance Sensors on the {\it Hubble Space Telescope}.  \citet{boden05} used the Keck Interferometer and published spectroscopic observations to determine the masses of the PMS binary HD 98800B.  Precise mass measurements of even a few systems provide critical data for calibrating the evolutionary tracks.  Additionally, the distances determined from the joint orbital solutions provide an accurate location of the system within the relatively spread-out star forming regions.

Haro 1-14c (HBC 644) is a PMS binary in the Ophiuchus star forming region.  \citet{reipurth02} first identified the system as a single-lined spectroscopic binary through observations in the visible part of the spectrum.  By observing in the infrared, where the light ratio between the primary and the cooler secondary is more favorable, Simon \& Prato (2004, SP04) measured the radial velocity amplitude of the secondary and determined a secondary to primary mass ratio of $0.310\pm0.014$.  In this paper we present spatially resolved observations of Haro 1-14c obtained with the Keck Interferometer.  We also present additional spectroscopic measurements to improve the precision of the velocity amplitude of the secondary.  These observations are described in \S~\ref{sect.obs}.  In \S~\ref{sect.orbfit}, we present results from a simultaneous orbit fit to the interferometric visibilities and the spectroscopic radial velocities.  In \S~\ref{sect.disc}, we discuss the derived distance to Haro 1-14c and compare the dynamical masses with predictions from evolutionary tracks.  A summary of the results appears in \S~\ref{sect.summ}.

\section{Observations of Haro 1-14c}
\label{sect.obs}

\subsection{Interferometric Observations}

The Keck Interferometer combines the light of the two 10-meter Keck telescopes and has a baseline of 85-meters, oriented 38$^\circ$ east of north \citep{colavita04, wizinowich04}.  The observable for the interferometer is the visibility or fringe contrast of the source.  For a binary star where the angular diameter of each component is unresolved, the normalized visibility $V^2$ at a wavelength $\lambda$ is
\begin{equation}
V^2 = \frac{1 + f^2 + 2 f \cos[{2\pi / \lambda} (u \Delta \alpha + v \Delta \delta)]}{(1 + f)^2}
\label{V2bin}
\end{equation}
where $(u,v)$ are the baseline components projected on the sky, $(\Delta\alpha,\Delta\delta)$ are the binary separations in RA and DEC, and $f$ is the component flux ratio (e.g. Boden 1999).  At the distance of the Ophiuchus star forming region, we expect the angular diameters of the stellar components of Haro 1-14c to be unresolved by the Keck Interferometer.    

We used the Keck Interferometer in its $V^2$ science mode, operating in the K-band.  We observed Haro 1-14c on five nights between June 2004 and May 2007.  To calibrate the fringe visibilities, we interspersed observations of unresolved single stars between those of Haro 1-14c.  An unresolved point source will have a visibility amplitude of 1.0.  We selected the calibrators using the getCal planning tool distributed by the Michelson Science Center (MSC).  The properties of Haro 1-14c and the selected calibrators are given in Table~\ref{tab.cal} which lists the RA and DEC coordinates, spectral type, V and K magnitudes, parallax, separation on the sky from Haro 1-14c, estimated angular diameter, and epochs of observation.  The angular diameters were calculated using the fbol routine in the getCal package which fits a blackbody model to spectrophotometric data.  Input to the routine included B and V magnitudes from the All-sky Compiled Catalog of 2.5 million stars (Kharchenko 2001), J, H, K band magnitudes from the Two Micron All-Sky Survey (2MASS) Catalog (Cutri et al. 2003), and parallaxes from the Hipparcos Catalog (Perryman et al. 1997).  The last three calibrators listed in Table~\ref{tab.cal} were observed as part of other programs but were used in this paper to improve the time sampling of calibrators surrounding the Haro 1-14c observations.

\begin{figure}
        \scalebox{0.40}{\includegraphics{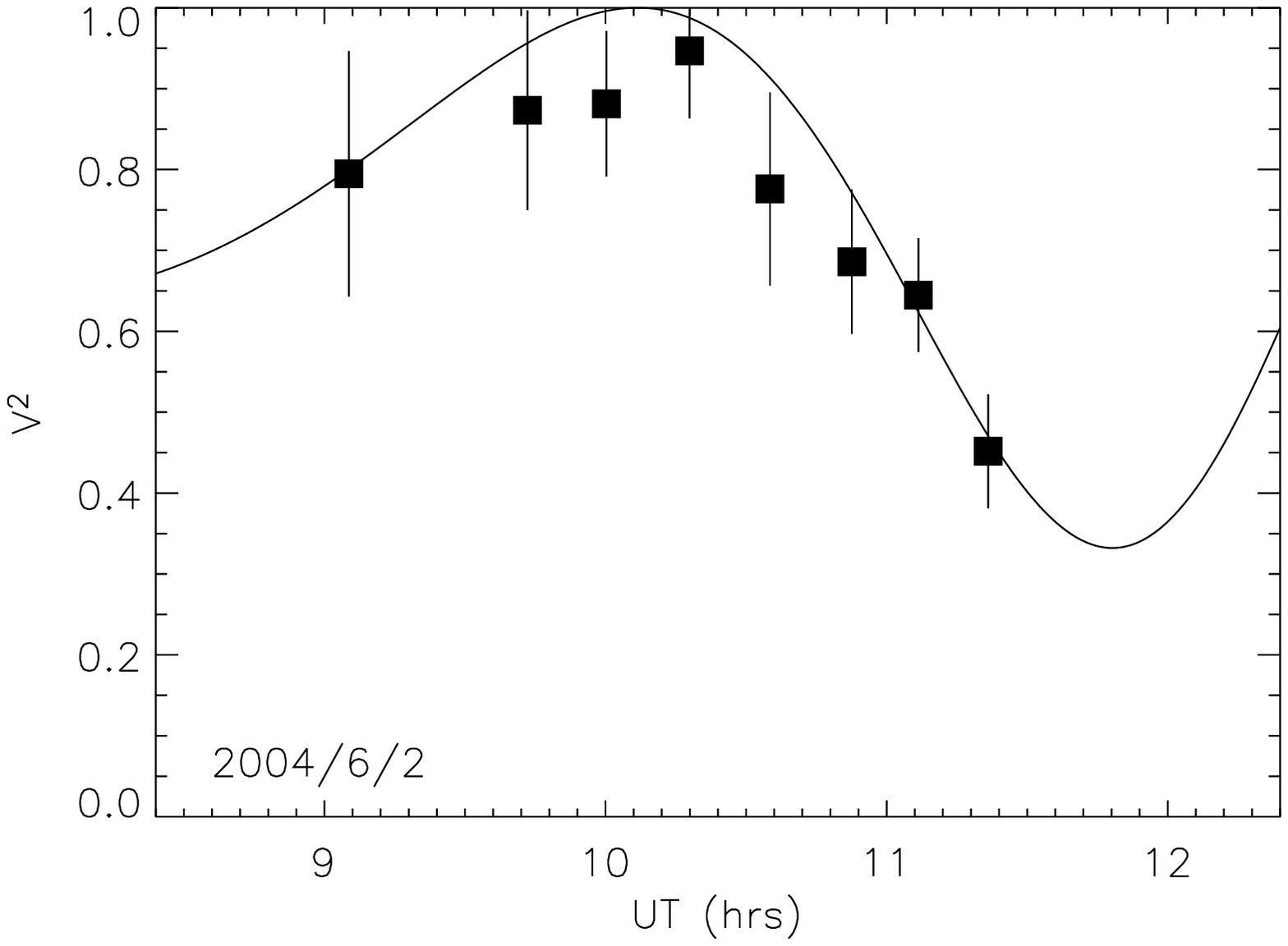}}
        \scalebox{0.40}{\includegraphics{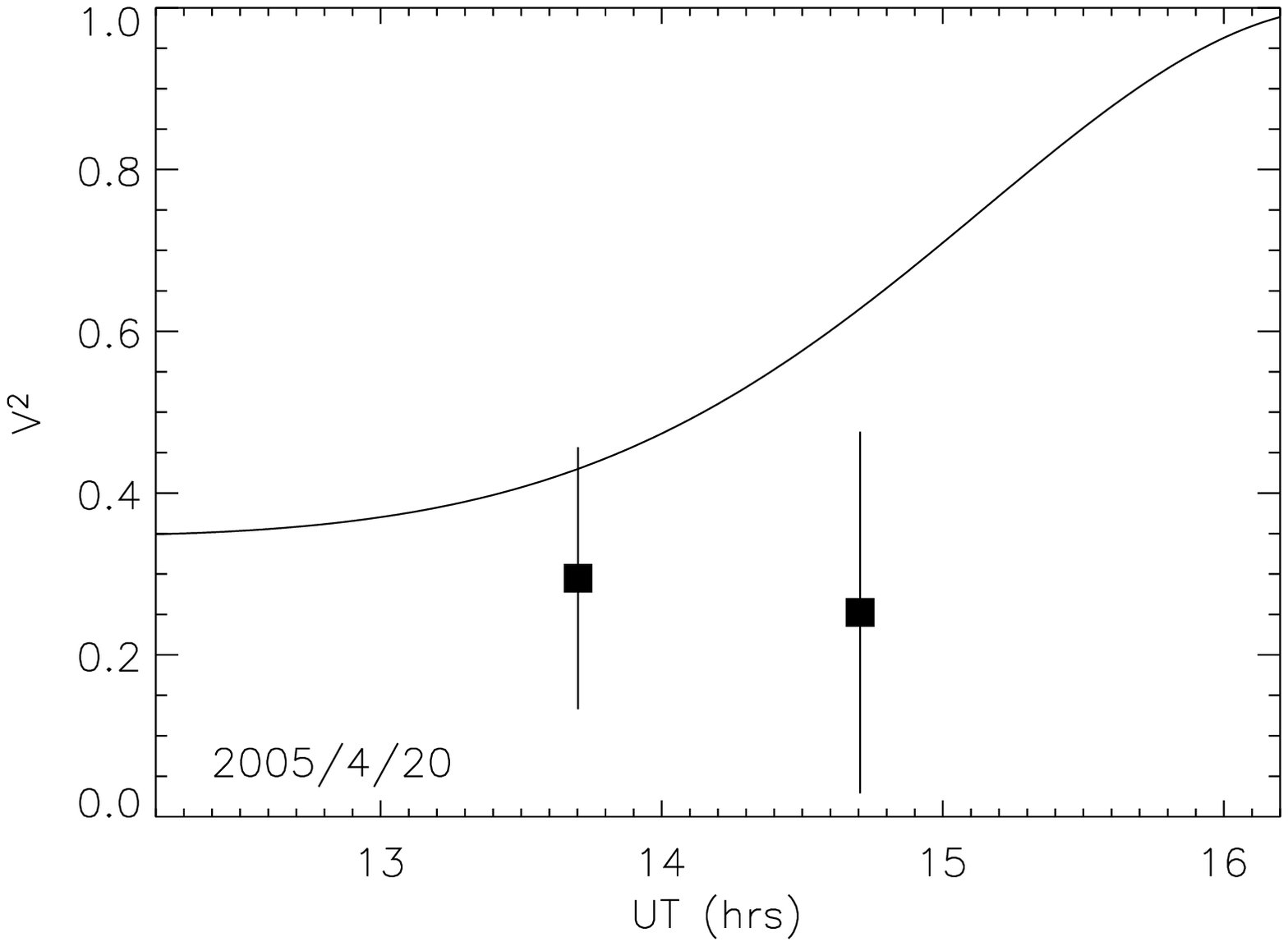}} \\
	\vspace{0.5cm}
        \scalebox{0.40}{\includegraphics{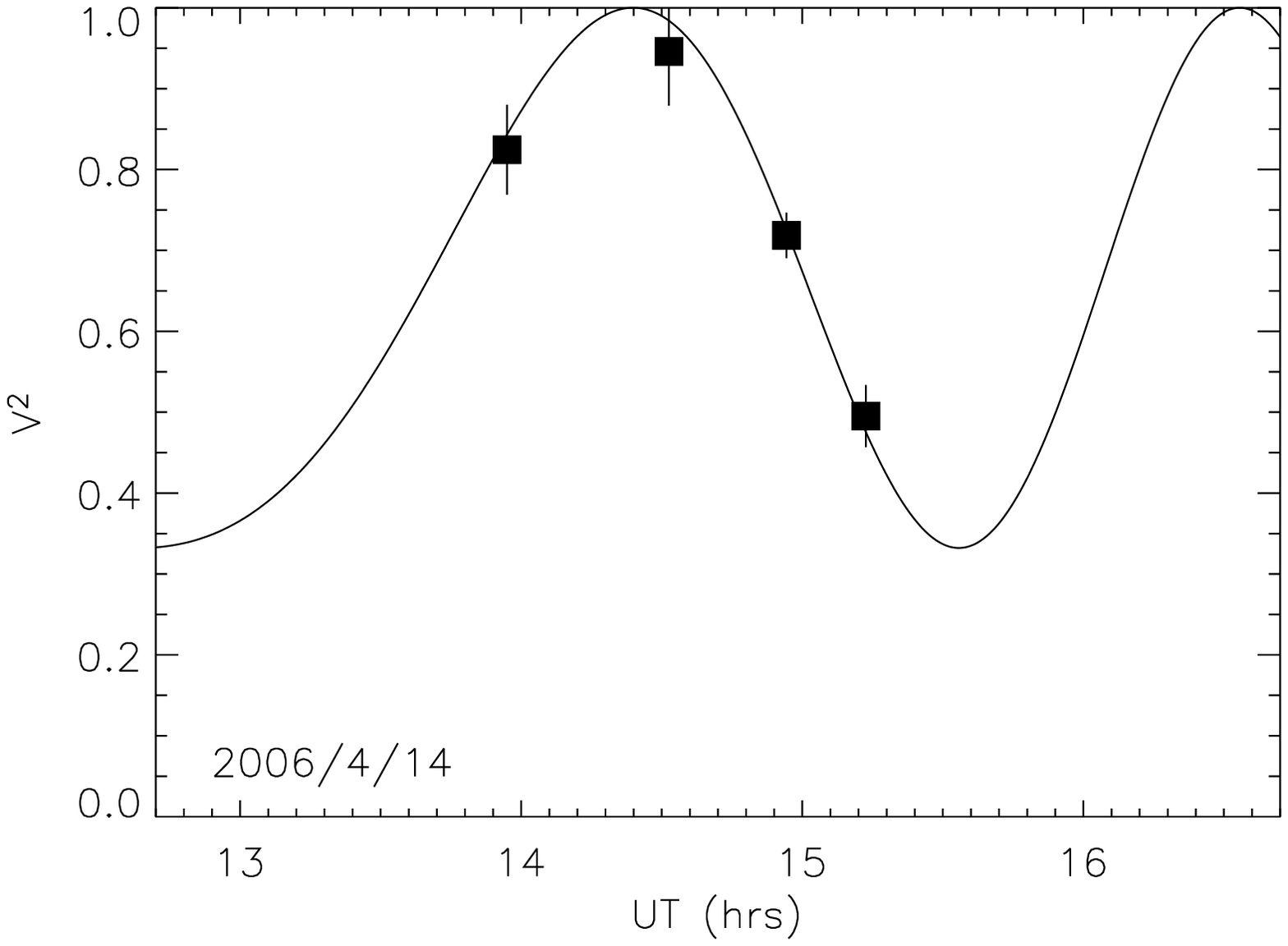}}
        \scalebox{0.40}{\includegraphics{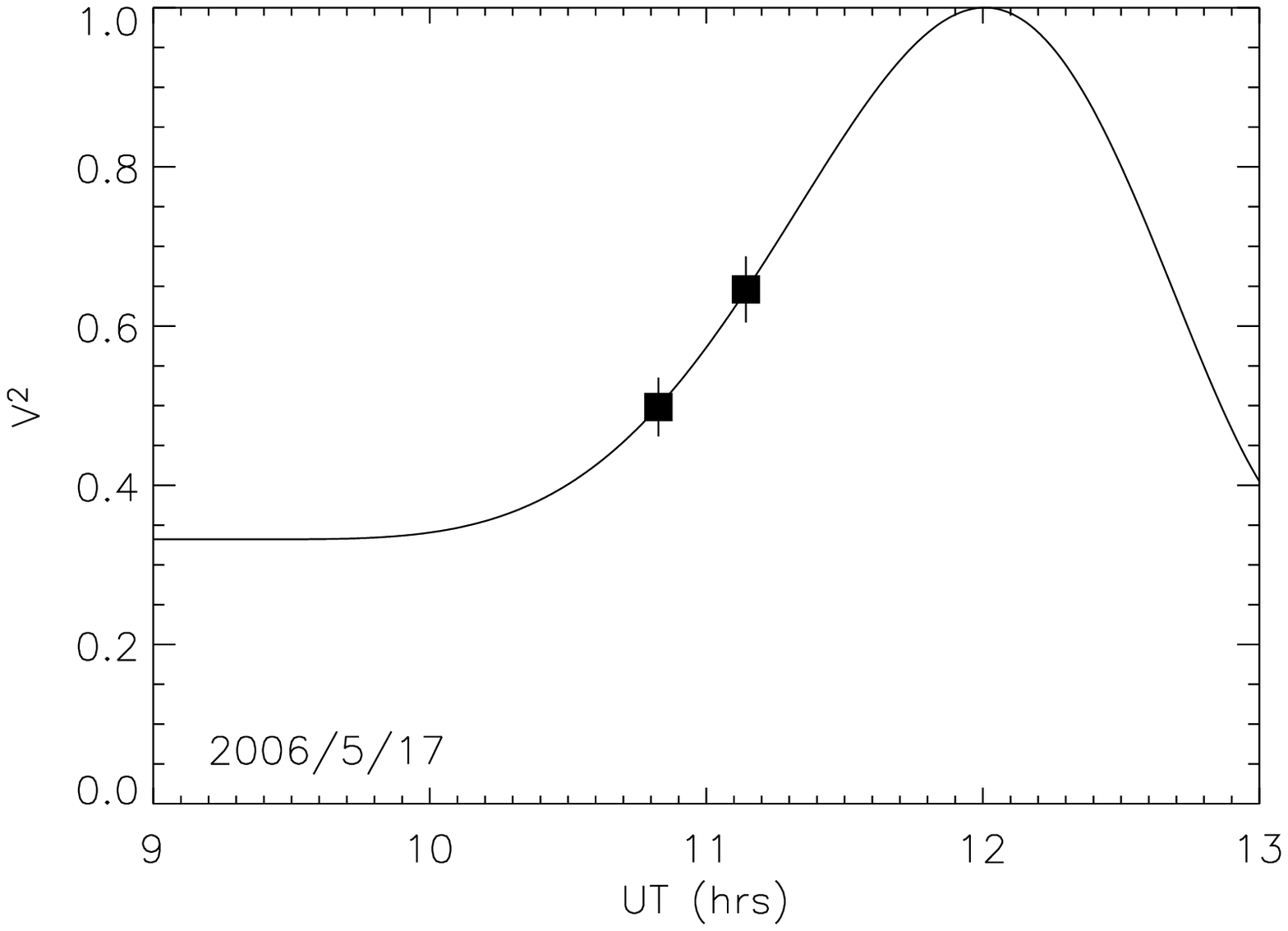}} \\
	\vspace{0.5cm}
        \scalebox{0.40}{\includegraphics{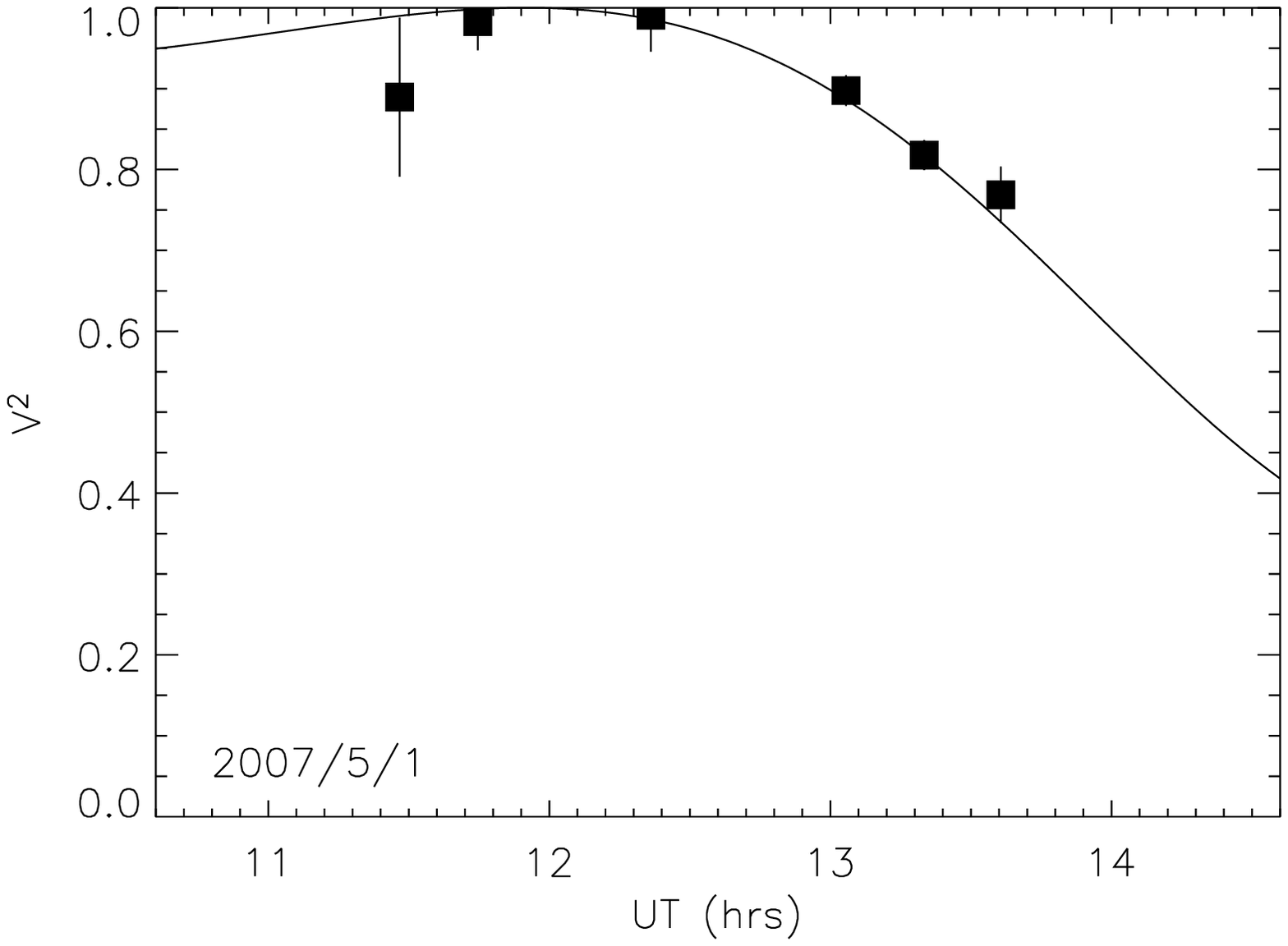}}
        \caption{Calibrated visibilities of Haro 1-14c measured with the Keck Interferometer on UT 2004 June 2, 2005 April 20, 2006 April 14, 2006 May 17, and 2007 May 1.  Overplotted in each panel is the best-fit binary orbit computed from a simultaneous fit to the $V^2$ measurements and the spectroscopic radial velocities.}
\label{fig.v2}
\end{figure}

\begin{figure}
	\scalebox{1.0}{\includegraphics{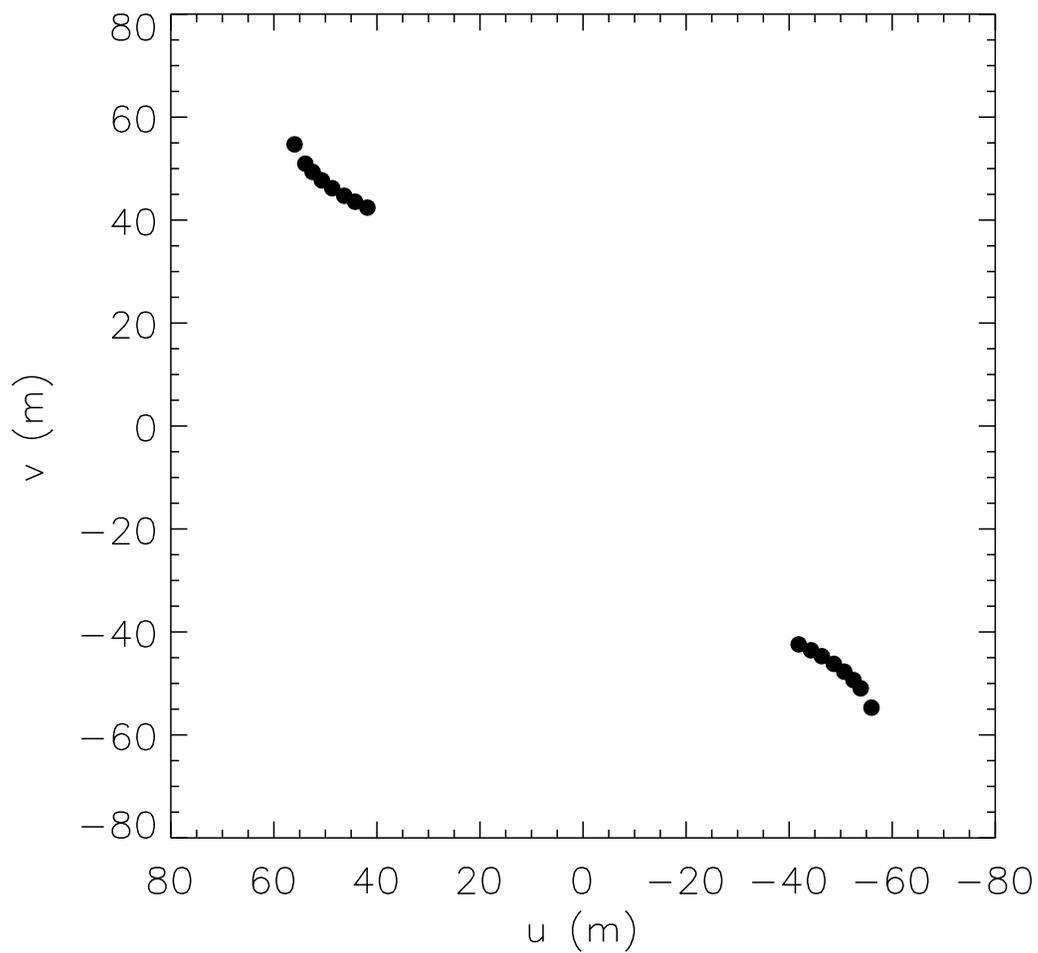}}
	\caption{$u$$v$-coverage sampled by the Keck Interferometer observations of Haro 1-14c on UT 2004 June 2.}
\label{fig.uv}
\end{figure}

We calibrated the raw visibilities using the wbcalib package developed by the MSC.  In doing so, we averaged the raw $V^2$ data over scan lengths of 300-500 sec.  We used the standard deviation of the raw measurements as an estimate of the uncertainties.  The wbcalib package computes the system visibility as a function of time based on the unresolved calibrator observations and applies this correction to the science data.  We also applied a correction to account for a flux-dependent bias between the target and the calibrators \citep{akeson07}.  Table~\ref{tab.v2} lists the Modified Julian Date (MJD), UT date and time, observed wavelength, calibrated $V^2$ and error, and the $u$ and $v$ projected baselines during the times of observations.  The calibrated visibilities of Haro 1-14c measured by the Keck Interferometer in 2004-2007 are plotted in Figure~\ref{fig.v2}.  Figure~\ref{fig.uv} shows an example of the $uv$-plane coverage obtained on 2004 June 2. 

\subsection{Velocity Measurements}

Near-infrared spectra were
obtained at the Keck II telescope using the NIRSPEC facility
spectrograph \citep{mclean98, mclean00}.  NIRSPEC employs
a 1024$\times$1024 ALADDIN InSb array detector with a plate
scale of 0.14$''$ in the dispersion direction and 0.19$''$
in the cross-dispersion direction.  The observations
were centered at 1.555 $\mu$m in the $H$-band.
A 0.288$''$ (2 pixel) by 24$''$ slit yielded high resolution
(R$=$30,000) spectra in several cross-dispersed orders. For the
new observations, 2$-$4 minute exposure times in nodded ABBA sets
were typical; the seeing was 0.6$-$0.8$''$.  Further information
on the previously published NIRSPEC data can be found in SP04.

We used only the central order, 49, in our analysis because it
covers a region rich in both atomic and molecular lines suitable
for the characterization of a range of spectral type components.
Furthermore, while this spectral region lacks contamination from
strong terrestrial absorption lines, more than 10 relatively
strong OH night sky emission lines span the order, providing
suitable dispersion calibration \citep{rousselot00}.
All data reduction was performed using the REDSPEC
package\footnote{http://www2.keck.hawaii.edu/inst/nirspec/redspec/index.html},
designed specifically for the reduction of NIRSPEC data.

Table 3 lists the UT date, Modified Julian Dates (MJD), velocities and
uncertainties for the primary and secondary, respectively, and the orbital
phase determined from the subsequent orbital fit (\S3.)  The first 6 lines
in Table 3 represent a reanalysis of the observations reported in SP04.
The last 2 lines refer to the new observations made after the publication of SP04.

To measure the spectral types and velocities of the components we analyzed 
all the spectra, including the spectra used in SP04, as 
follows.  We first used TODCOR, the two-dimensional cross-correlation 
algorithm written for analysis of double-lined spectroscopic binaries by 
\citet{zucker94} and our suite of observed stellar spectral templates 
\citep{prato02} to  identify the spectral templates providing the 
best match in spectral type and rotational broadening. HR 8085, a K5~V, 
and GL 15A, a M1.5~V, both rotationally broadened to 12 km s$^{-1}$, 
provided the best matches for the primary and secondary of Haro 1-14c, 
consistent with the result reported in SP04.

We chose not to use the observed stellar templates to measure the component velocities, as we had in SP04, because to do so would have propagated velocity uncertainties associated with the templates to Haro 1-14c.  Instead, we used a suite of theoretical spectra calculated from updates of the NextGen models \citep{hauschildt99}.  These calculations represent main sequence stars with solar abundance in the mass range 0.1 to 1.6 $M_{\odot}$.  We selected theoretical spectra that provided excellent matches to the observed spectra of HR 8085 and GL 15A.  We then used these two corresponding spectra, with a velocity correction to the heliocentric frame appropriate for each target observation, to measure the radial velocities of the primary and secondary  components of Haro 1-14c.  We applied Wilson's (1941) approach to our 8 pairs of primary  and secondary velocities to determine  the center-of-mass velocity and its uncertainty, $\gamma = -9.38 \pm 0.39$ km s$^{-1}$.  This value of $\gamma$ and the component velocities we measured from the infrared spectra are in the laboratory frame of the theoretical spectra, which is not necessarily the frame in which \citet{reipurth02} measured the velocity of the Haro 1-14c  primary.  Their 56 measurements yield a very accurate value of the center-of-mass velocity of $\gamma = -8.71 \pm0.07$ km s$^{-1}$.  We chose therefore to shift our velocity measurements by the difference between our and Reipurth et al.'s value of $\gamma$, $-0.67 \pm 0.39$ km s$^{-1}$, effectively transforming them into the frame of Reipurth et al. These values, and their propagated uncertainties, are reported in Table 3.    The velocity uncertainties of the secondary, which are particularly important in the determination of the secondary/primary mass ratio, are about half the values reported in SP04.

\section{Orbital Fit to Haro 1-14c}
\label{sect.orbfit}

Velocity measurements of a double-lined spectroscopic binary yield  the orbital period, eccentricity, time of periastron passage, longitude of periastron, and the velocity semi-amplitudes  of the primary and secondary ($P$, $T$, $e$, $\omega$, $K_1$, and $K_2$).  A visual binary yields $P$, $T$, $e$, $\omega$, $a('')$, $i$, and $\Omega$, where the last three parameters are the semi-major axis in arcseconds, the inclination, and the position angle of the line of nodes.  Combining the visual and spectroscopic orbital parameters gives the masses of the primary and secondary,
\begin{equation}
M_1 = {\frac{1.036\times 10^{-7} (K_1 +K_2)^2 K_2 P (1-e^2)^{3 / 2}}{sin^3 i}}
\label{eq.M1}
\end{equation}
\begin{equation}
M_2= {{1.036\times 10^{-7} (K_1 +K_2)^2 K_1 P (1-e^2)^{3 / 2} } 
\over {sin^3 i}}
\label{eq.M2}
\end{equation}
where $K_1$ and $K_2$ are in km~s$^{-1}$, $P$ in days, and $M_1$ and $M_2$ in $M_{\odot}$ \citep{fernie00}.  The distance to the binary is determined by comparing the angular size of the visual orbit with the physical scale of the spectroscopic orbit,
\begin{equation}
d = \frac{9.198\times 10^{-5} (K_1+K_2) P (1- e^2)^{1 / 2}}{a('')\sin{i}}
\label{eq.dist}
\end{equation}
where the distance $d$ is in pc.

\subsection{Formal Orbital Analysis}

We performed a simultaneous orbit fit to the interferometer visibilities and spectroscopic radial velocities presented in this paper and in \citet{reipurth02}.  The orbit fitting procedure starts with initial estimates of the orbital parameters ($P$, $T$, $e$, $\omega$, $a$, $i$, $\Omega$, $K_1$, $K_2$, $\gamma$) and the K-band flux ratio and follows a Newton-Raphson approach to adjust these parameters to their best-fit values.  During each iteration of the routine, the values of ($\Delta\alpha$, $\Delta\delta$) are computed at the times of the interferometer observations from the given set of orbital parameters.  The program then calculates model visibilities according to Equation \ref{V2bin}, assuming that the angular diameters of the components are unresolved.  The model radial velocities are determined from
\begin{equation}
v_1 = +K_1[e\cos{\omega} + \cos{(\nu + \omega)}] + \gamma
\label{vel1}
\end{equation}
\begin{equation}
v_2 = -K_2[e\cos{\omega} + \cos{(\nu + \omega)}] + \gamma
\label{vel2}    
\end{equation}
where the true anomaly $\nu$ is computed at the times of the spectroscopic observations from the given set of orbital parameters.
The program adjusts the initial parameters to minimize the $\chi^2$ between the measurements and model values, which is given by
\begin{equation}
\chi^2 = \sum \frac{(V^2_{\rm meas} - V^2_{\rm fit})^2}{\sigma_{V^2}^2} +
	 \sum \frac{(v_{1\rm meas} - v_{1\rm fit})^2}{\sigma_{v_1}^2} + 
	 \sum \frac{(v_{2\rm meas} - v_{2\rm fit})^2}{\sigma_{v_2}^2}
\end{equation}
where the 'meas' and 'fit' subscripts refer to the measured values and the model fit, respectively.

\begin{figure}
\begin{center}
\scalebox{1.0}{\includegraphics{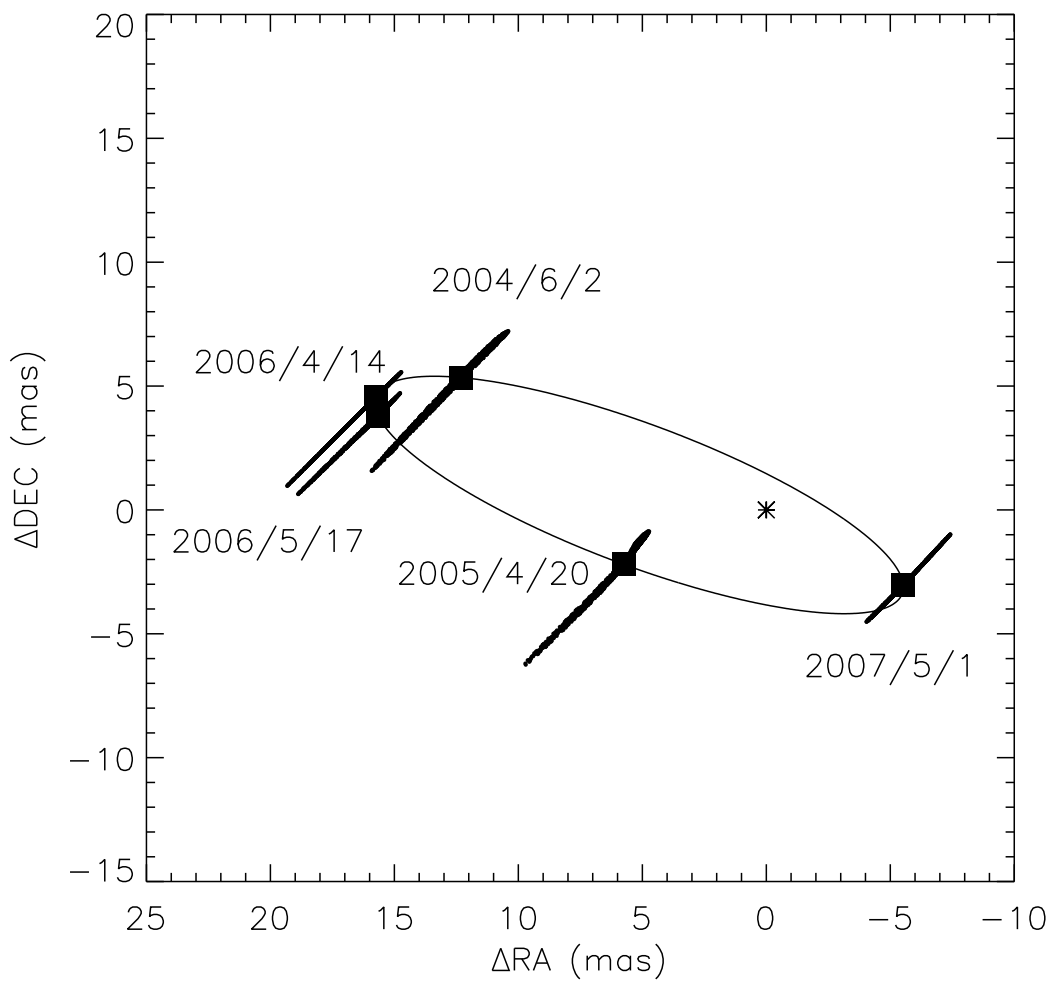}}
\caption{Best-fit orbit for Haro 1-14c.  The large squares indicate the orbital position at the epochs of the Keck Interferometer observations.  The elongated error ellipses represent the 1-$\sigma$ confidence intervals ($\Delta\chi^2 = 1$) determined from the Monte Carlo search of the orbital parameter space.  The narrowness of the ellipses (0.1-0.3 mas) reflects the precision along the direction of the interferometer baseline.}
\label{fig.orb}
\end{center}
\end{figure}

\begin{figure}
\begin{center}
        \scalebox{0.80}{\includegraphics{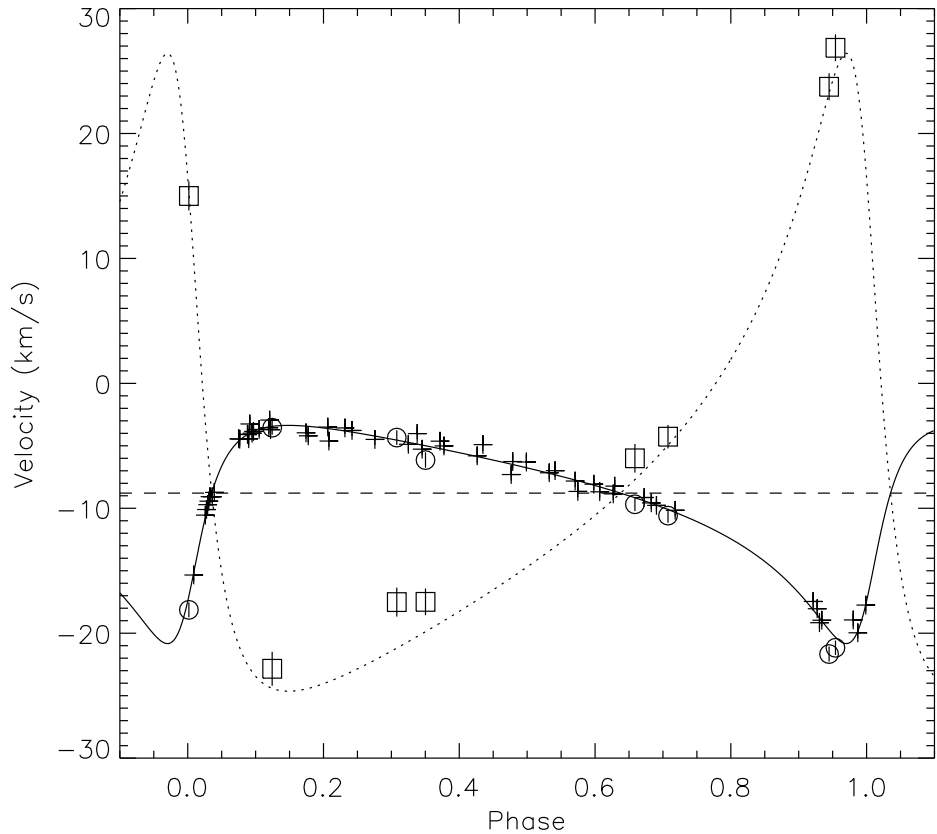}}
        \caption{Radial velocity as a function of phase for Haro 1-14c.  The crosses represent the Reipurth et al. (2002) measurements of the primary velocity.  The open symbols and $1 \sigma$ uncertainties indicate the radial velocities for the primary (circles) and secondary (squares) measured using NIRSPEC.  The solid line shows the best fit to the primary star data, and the dotted line for the secondary.  The horizontal dashed line indicates the center-of-mass velocity of the system.  The abscissa spans a 20\% redundancy in phase for clarity.}
\label{fig.sb2}
\end{center}
\end{figure}

Table~\ref{tab.orbpar} lists the orbital parameters, component masses, and distance determined from the simultaneous orbit fit.  The last two rows show the $\chi^2$ of the fit and the reduced $\chi^2_\nu$ for 83 degrees of freedom.  The best-fit orbit is overplotted in the $V^2$ curves shown in Figure~\ref{fig.v2}.  Figure~\ref{fig.orb} shows the orientation of the binary on the plane of the sky.  The epochs of the Keck Interferometer observations are marked along the orbit.  The radial velocity measurements and best-fit velocity curves are shown in Figure~\ref{fig.sb2}.

\begin{figure}
        \scalebox{1.0}{\includegraphics{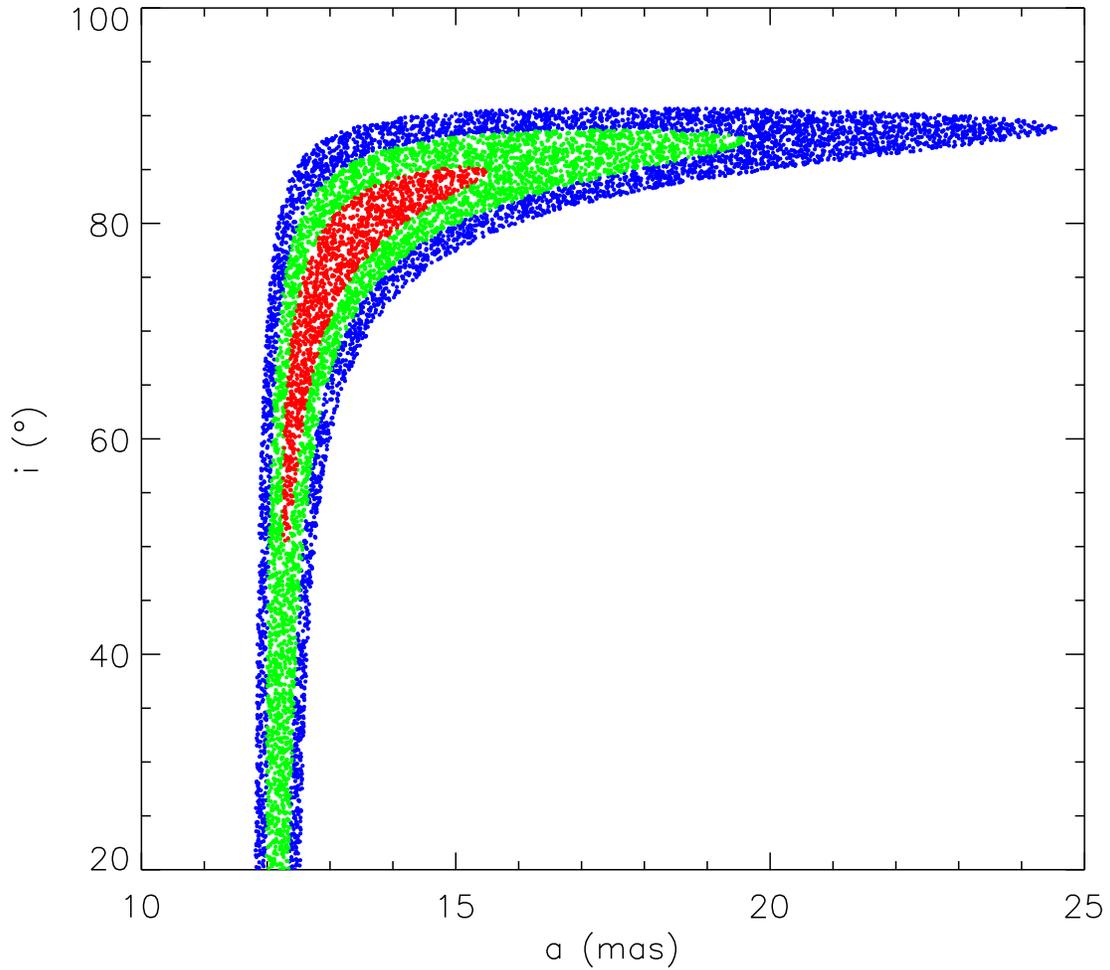}}
        \caption{$\chi^2$ surfaces determined through the two-dimensional Monte Carlo search for orbital solutions that fit the current data set.  The color codes correspond to the 1~$\sigma$ (red), 2~$\sigma$ (green), and 3~$\sigma$ (blue) confidence intervals ($\Delta\chi^2$ = 1, 4, and 9, respectively).}
\label{fig.chi_ai}
\end{figure}

The spectroscopic orbital parameters ($P$, $T$, $e$, $\omega$, $K_1$, $K_2$, $\gamma$) of Haro 1-14c are well-determined by the single- and double-lined radial velocities.  However, the limited sampling of the visual orbit provided by the Keck Interferometer observations produces a large uncertainty in the inclination.  We examined the distributions of $a$ and $i$ allowed by the current set of $V^2$ measurements by performing a two-dimensional Monte Carlo search.  During each iteration of the search, we selected values of $a$ and $i$ at random and minimized the remaining orbital parameters through the standard Newton-Raphson approach.  We accumulated a total of 10,000 possible solutions within the 3-$\sigma$ confidence interval ($\Delta\chi^2 = 9$ from the minimum $\chi^2$ value).  The resulting $\chi^2$ distribution is shown in Figure~\ref{fig.chi_ai}.    The long tail of solutions that extends to small inclinations produces extremely large component masses when combined with the spectroscopic parameters.  The uncertainties for $a$, $i$, $\Omega$, $M_1$, $M_2$, and $d$ quoted in Table~\ref{tab.orbpar} represent the bounds of the 1-$\sigma$ ($\Delta\chi^2 = 1$) confidence regions.

The results from the Monte Carlo search also allow us to examine the uncertainties in the binary separation derived from the epochs of the Keck Interferometer observations.  For each of the 10,000 orbital solutions found, we computed the binary separations at the times of the interferometer observations and overplotted the 1-$\sigma$ confidence intervals in the orbit plot in Figure~\ref{fig.orb}.  The elongated error ellipses arise from the limited $uv$-coverage sampled during the interferometer observations (see Figure~\ref{fig.uv}).  Given the low altitude of Haro 1-14c at the latitude of the Keck Observatory, the short 1-3 hour observing sessions do not provide much rotation of the projected interferometer baseline on the sky.  Essentially, the narrowness of the error ellipses shows the exceptional resolution along the direction of the interferometer baseline.  Because we fit all of the visibilities from each epoch simultaneously with the spectroscopic radial velocities, additional measurements with the Keck Interferometer at different locations along the orbit will help reduce the range of allowed orbital solutions and hence reduce the size of the error ellipses in the binary separation for every epoch (see \S~\ref{sect.sim} for more details). 

\subsection{Masses and Distance from a Statistical Analysis of Orbital Solutions}
\label{sect.stat}

To investigate the distribution of masses and distance allowed by the interferometric and spectroscopic data, we performed a more exhaustive 6-dimensional Monte Carlo search.  We randomly selected values for all of the orbital parameters on which the masses and distance depend ($P$, $e$, $a$, $i$, $K_1$, and $K_2$) while minimizing the fit to the remaining parameters through the standard Newton-Raphson approach.  We accumulated a total of 10,000 possible solutions within the $\Delta\chi^2 = 9$ confidence interval.  

\begin{figure}
\begin{center}
        \scalebox{0.33}{\includegraphics{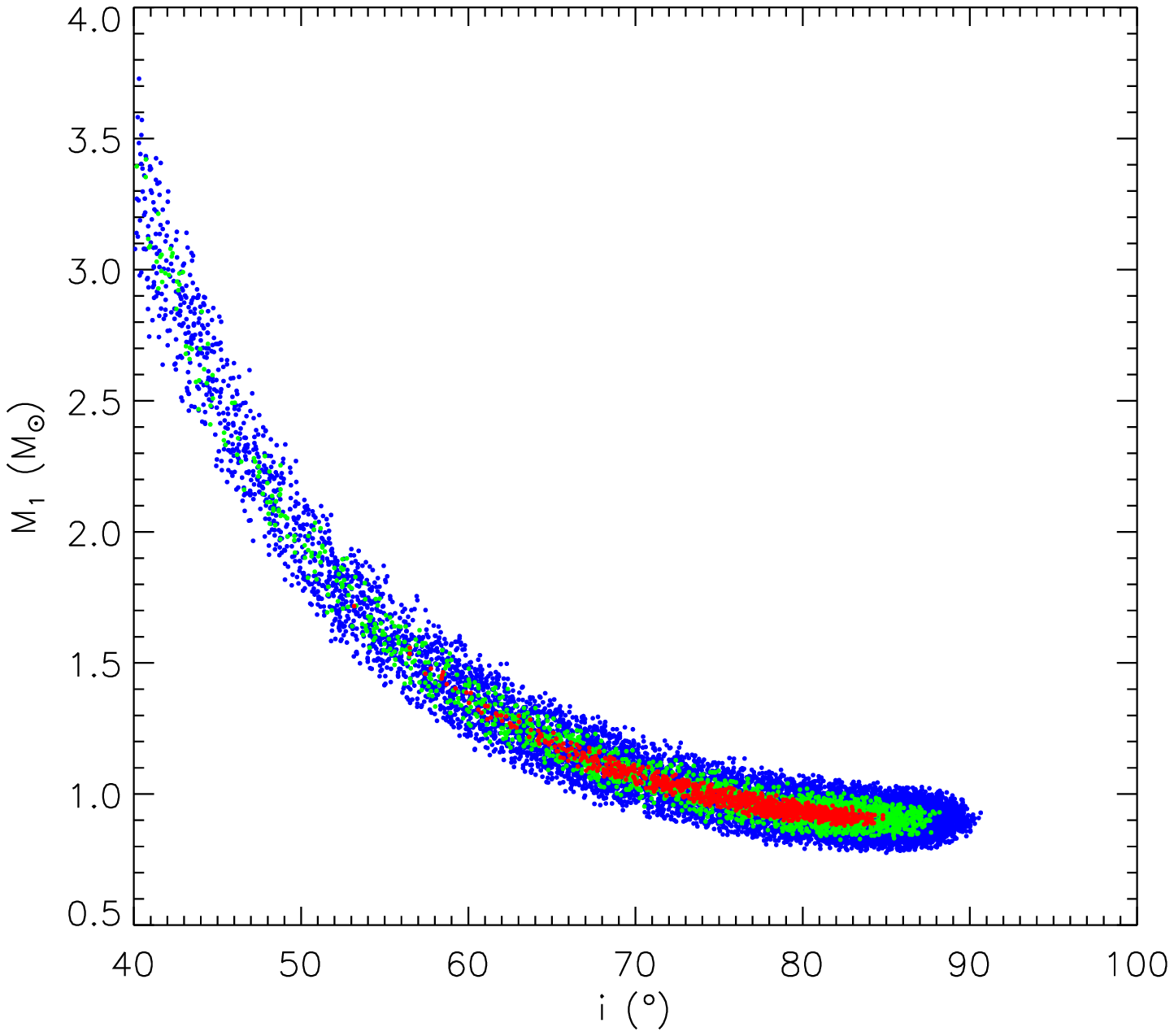}}
        \scalebox{0.33}{\includegraphics{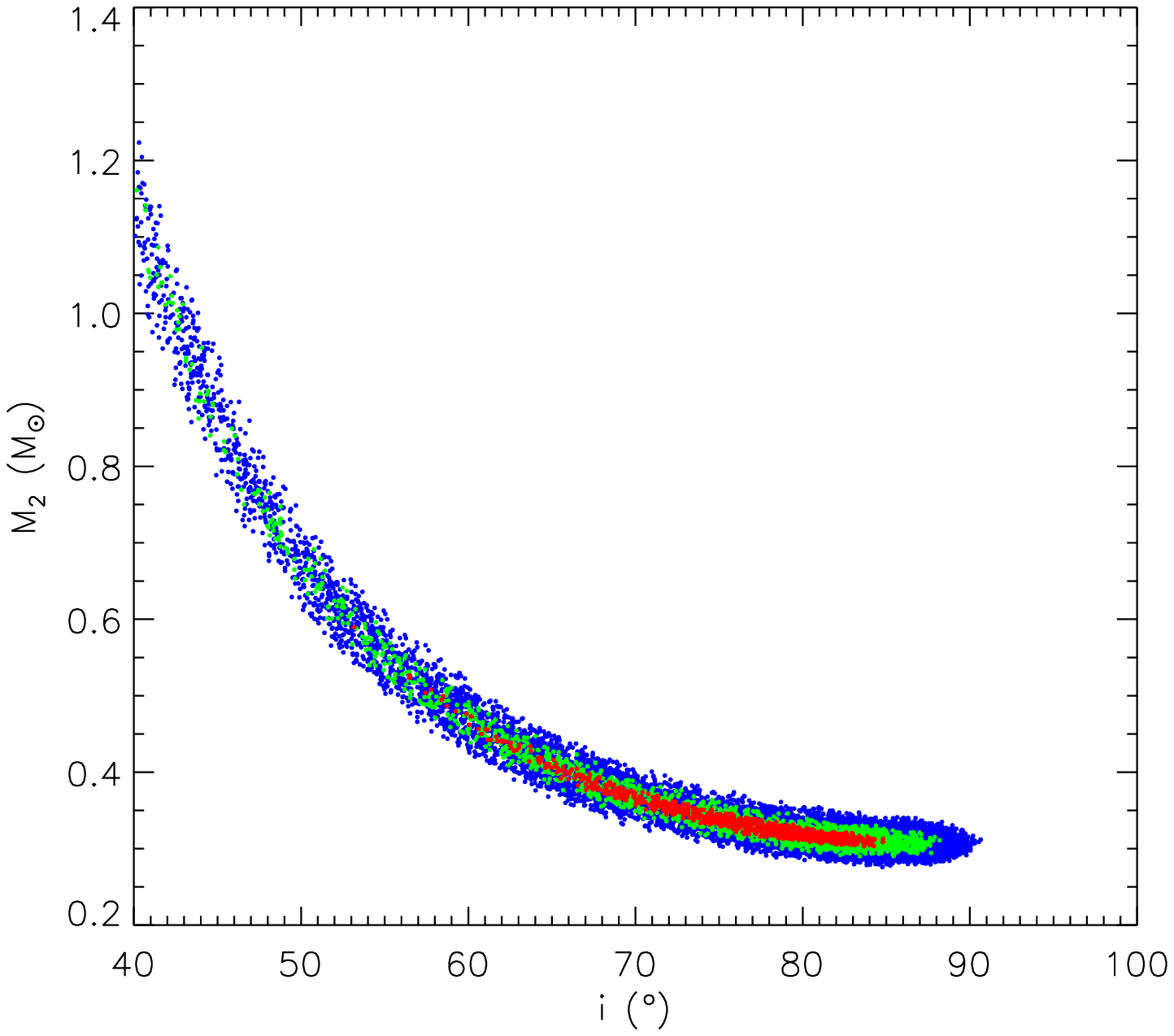}} 
        \scalebox{0.33}{\includegraphics{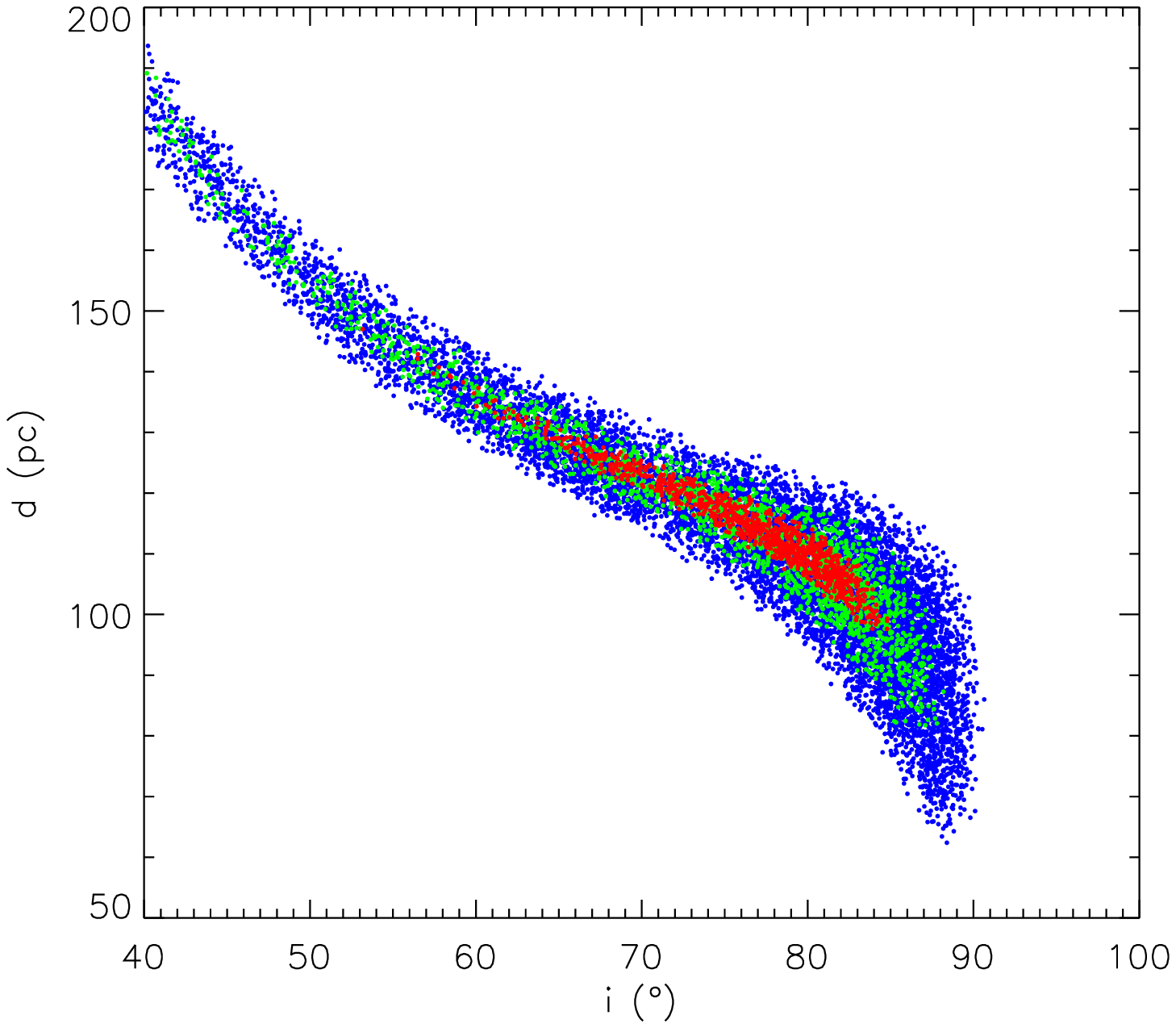}} \\
        \scalebox{0.30}{\includegraphics{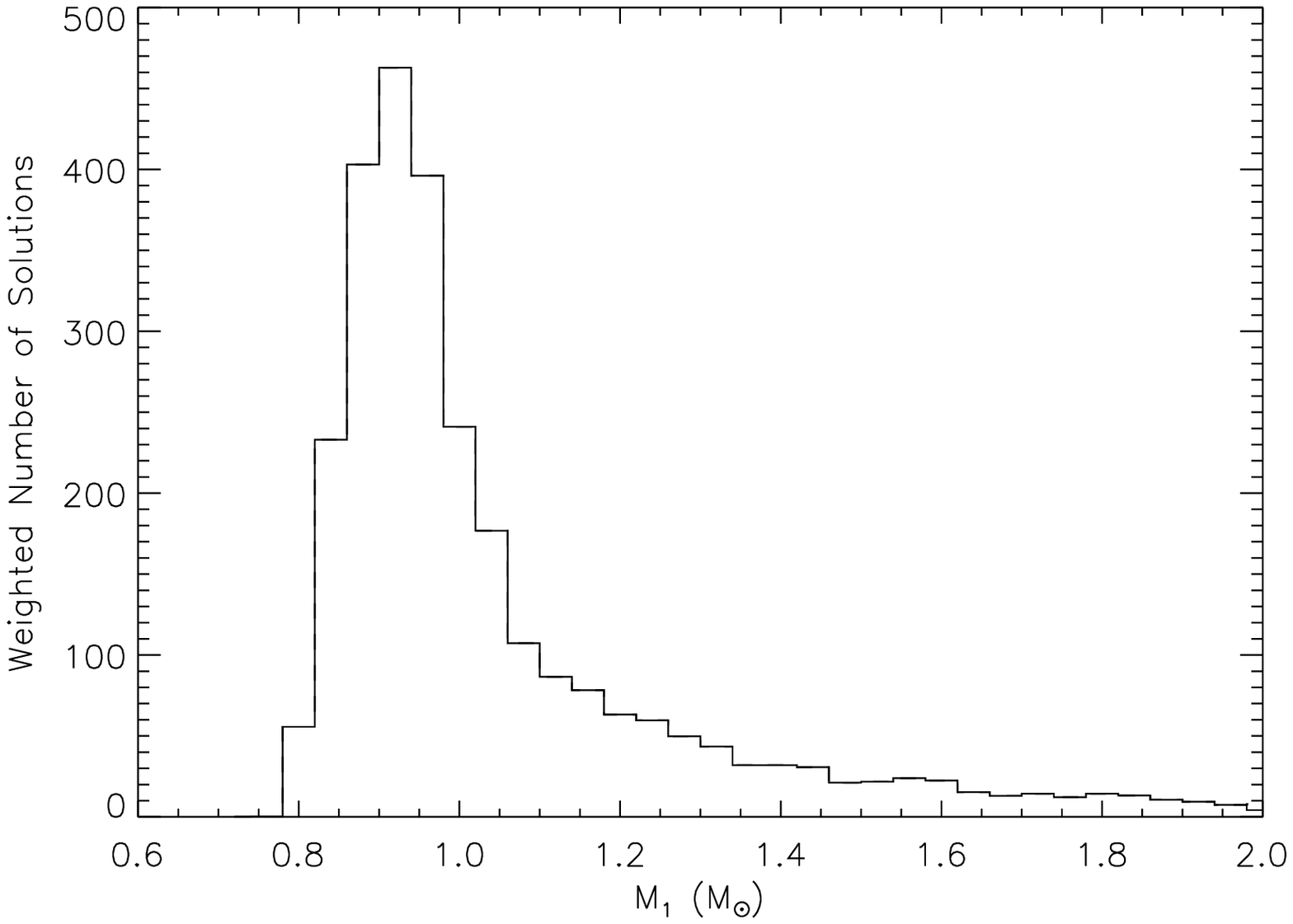}}
        \scalebox{0.30}{\includegraphics{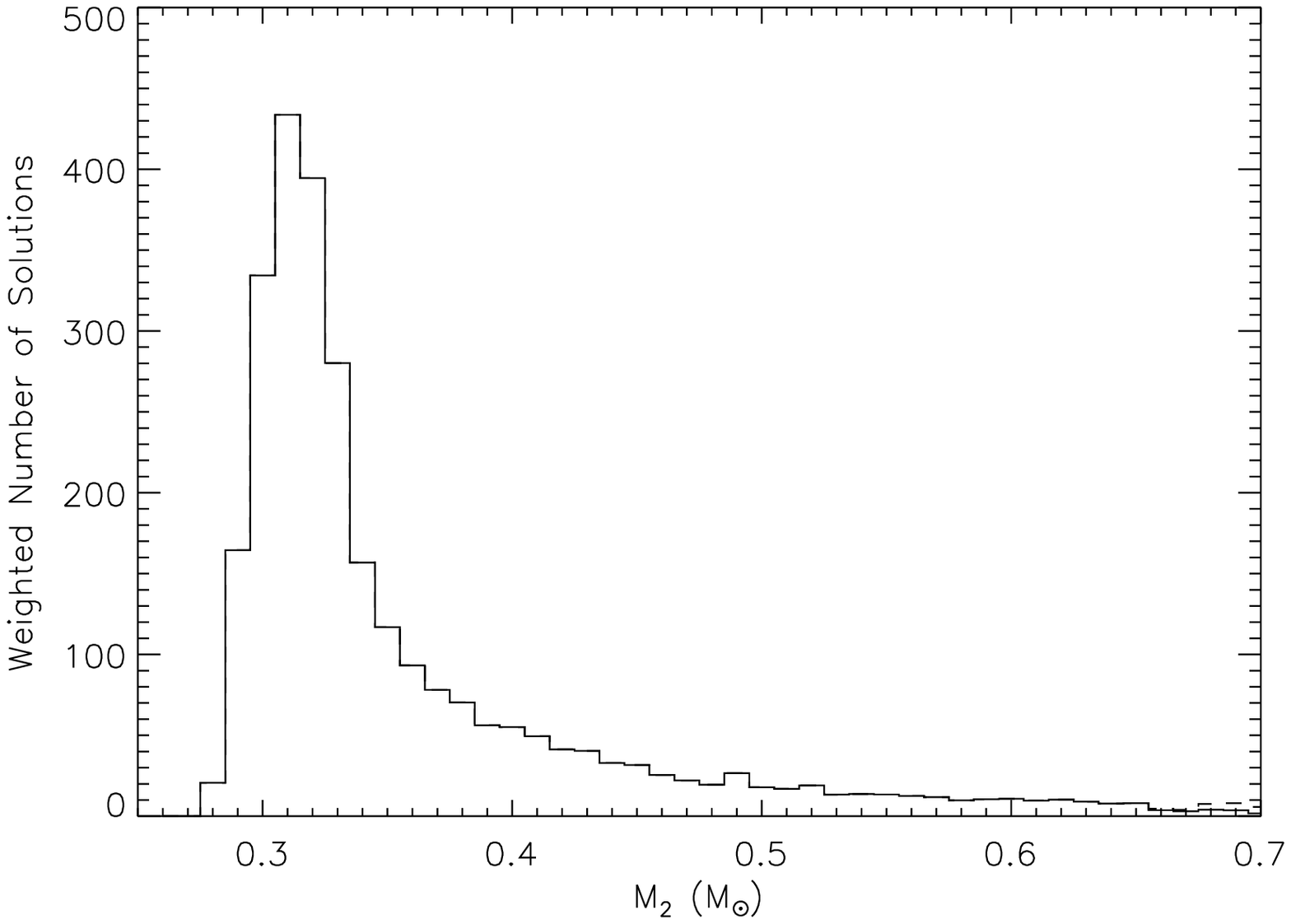}}
        \scalebox{0.30}{\includegraphics{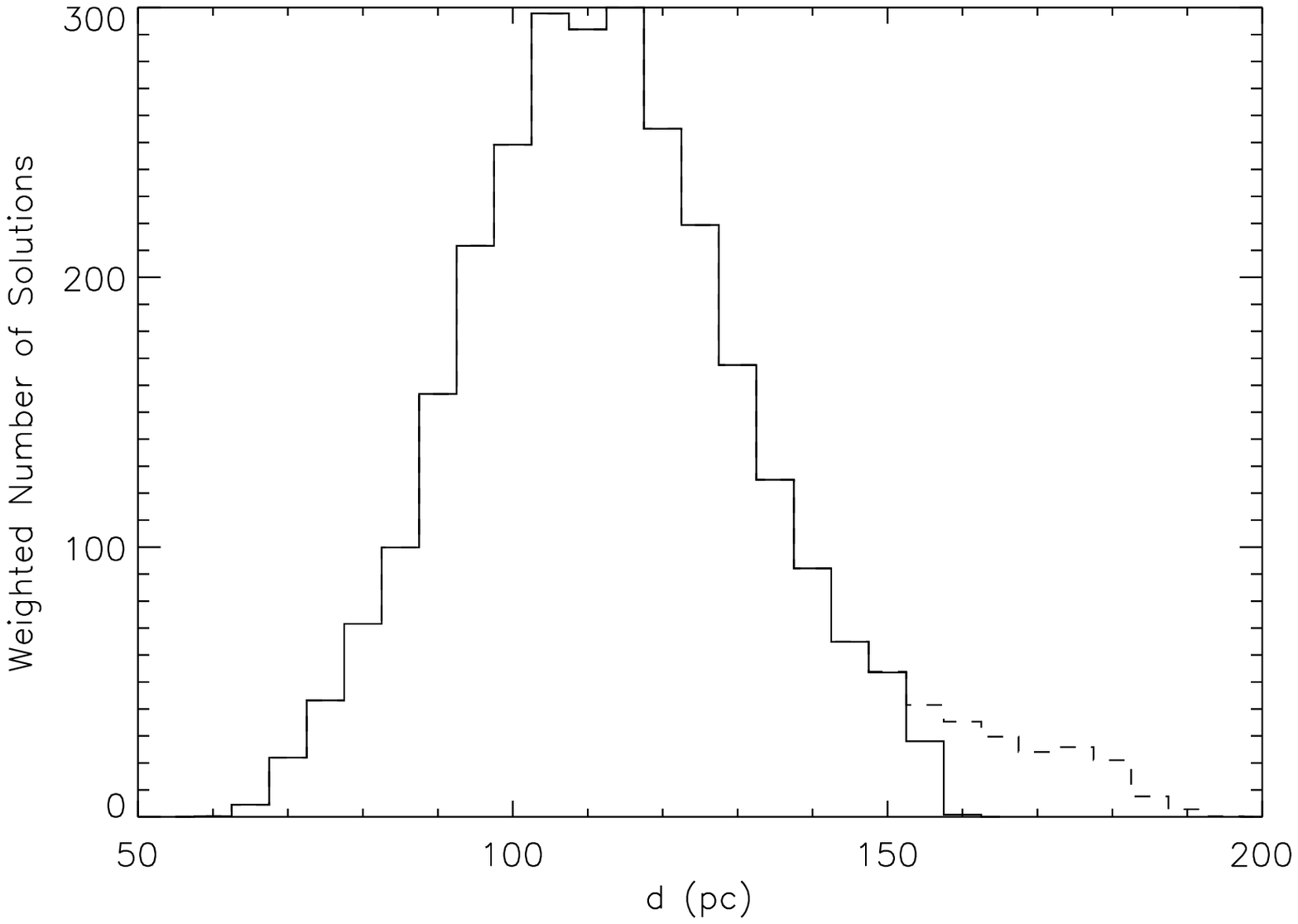}}
\end{center}
        \caption{{\it Top row:} Crosscuts through the $\chi^2$ surface for $M_1$ (left), $M_2$ (middle), and distance (right) versus the inclination based on the results from the 6-dimensional Monte Carlo search described in \S~\ref{sect.stat}.  The color codes correspond to the 1~$\sigma$ (red), 2~$\sigma$ (green), and 3~$\sigma$ (blue) uncertainty intervals ($\Delta\chi^2$ = 1, 4, and 9, respectively).  We added an additional 1,000 solutions in each of the 1~$\sigma$ and 2~$\sigma$ intervals to better define these regions visually in the plots; these added solutions were not used in the statistical analysis of the distributions.  {\it Bottom row:} Weighted distributions for $M_1$ (left), $M_2$ (middle), and distance (right).  In constructing the histograms, each orbital solution was weighted by its $\chi^2$ probability.  The dashed line indicates the solutions removed by the 2~$M_\odot$ upper mass cut-off for the primary.}
\label{fig.mass}
\end{figure}

The top row of Figure~\ref{fig.mass} shows cross-cuts through the $\chi^2$ surface for the component masses and distance versus the inclination.  Following the approach outlined in \citet{schaefer06}, we constructed histograms of the allowed masses and distance where each solution is weighted by its $\chi^2$ probability.  As seen in these histograms, shown in the bottom row of Figure~\ref{fig.mass}, the mass and distance distributions are strongly peaked.  We present the median values of these distributions in Table~\ref{tab.mass}.  The range of the quoted uncertainties include 34\% of the values on each side of the median.  In computing these values, we applied an arbitrary upper mass cut-off of 2.0~$M_\odot$ for the primary, since masses larger than this value are clearly not representative of the spectral type and stellar luminosity of Haro 1-14c.  Compared to the formal $\chi^2$ intervals in Table~\ref{tab.orbpar}, the statistical analysis substantially reduces the size of the error bars.  It is worth noting that because of the sharp drop in the mass distributions on the low-mass sides, improved mass values are not likely to be much smaller than the present values, but may be larger.  In the following sections we refer to the median masses and distance estimates given in Table~\ref{tab.mass}, as we believe these values more accurately represent the asymmetric parameter distributions.

\subsection{Comments on the Reliability of the Orbital Fit and Expectations for the Future}
\label{sect.sim}

Obtaining interferometric observations that are well distributed across the orbit is critical for determining a reliable measurement of the inclination.  Since the inclination is determined from the offset of the primary star from the focal point of the apparent orbit, measuring an accurate value requires good orbital sampling along both the major and minor axes of the apparent ellipse.  For the orbit of Haro 1-14c, we have fairly good coverage along the major axis of the ellipse, but we have only one marginal measurement along the direction of the minor axis.  During the observations on 2005 April 20, we experienced unusually low instrument throughput on Haro 1-14c which made it difficult to lock onto the fringes and consequently degraded the quality of the $V^2$ measurements.  Because these marginal measurements sample a critical location along the orbit, a precise value of the inclination has not yet been determined.  

\begin{figure}
        \scalebox{1.0}{\includegraphics{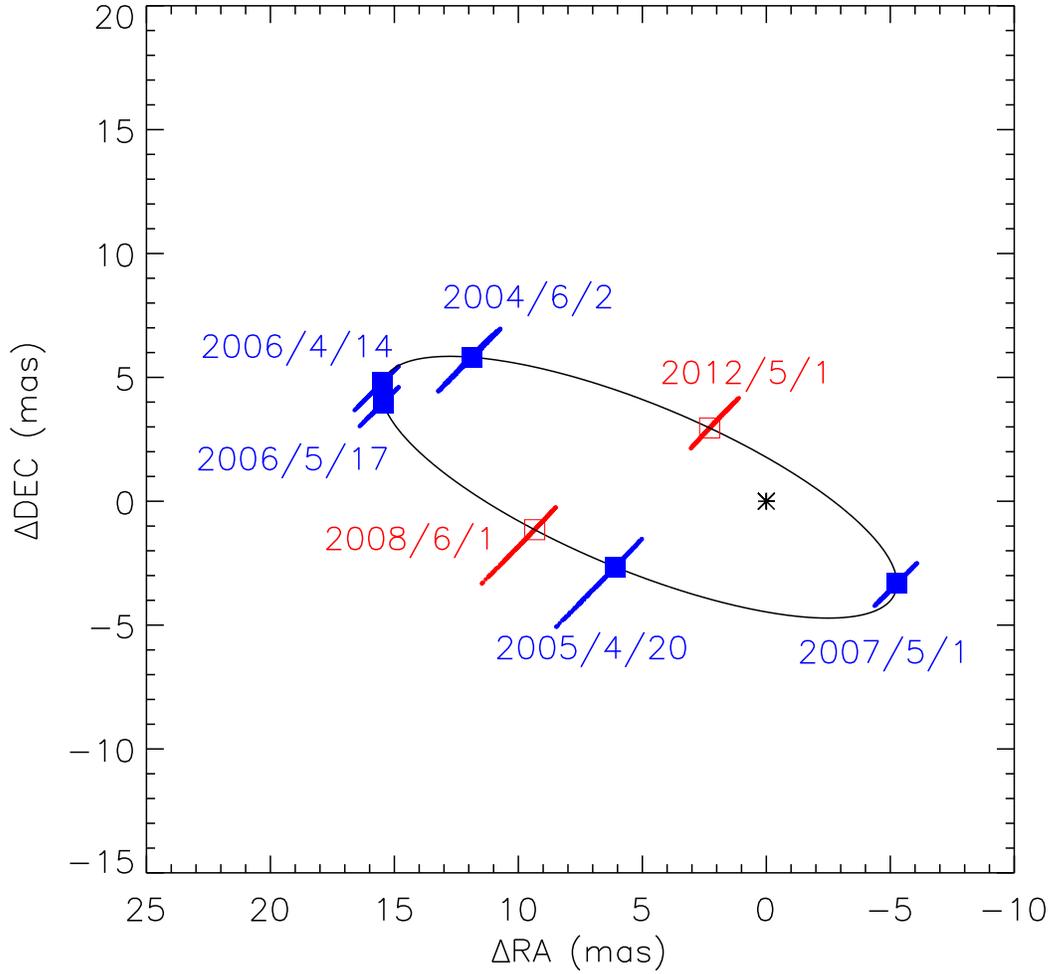}}
        \caption{Modeled orbit fit based on simulated Keck Interferometer data for 2008 June 1 and 2012 May 1.  These dates were selected to provide good sampling of the Haro 1-14c orbit at times when the binary is observable with the Keck Interferometer.  The expected improvement in the orbital fit with additional measurements is clearly apparent when comparing to the current fit in Figure~\ref{fig.orb}.}
\label{fig.sim}
\end{figure}

To investigate what we can expect from future observations with the Keck Interferometer, we simulated $V^2$ measurements for 2008 June 1 and 2012 May 1.  These dates were selected to provide good sampling of the Haro 1-14c orbit at times when the binary is observable with the Keck Interferometer.  To simulate the interferometer measurements, we computed the expected $V^2$ based on the predicted binary position at 20 minute intervals over 2.3 hours centered on transit.  We added Gaussian noise to the simulated $V^2$ and recomputed the simultaneous orbit fit based on the measured and simulated $V^2$ and spectroscopic radial velocities.  As seen in Figure~\ref{fig.sim}, the additional measurements would significantly reduce the uncertainties in the orbit fit.  Moreover, the simulated data yield formal $\chi^2$ uncertainties in the masses and distance that are smaller by more than a factor of 2 compared with the current values in Table~\ref{tab.orbpar}.

\section{Discussion}
\label{sect.disc}

\subsection{Distance to Haro 1-14c}
\label{sect.dist}

The orbital measurements yield a dynamical estimate of the distance to Haro 1-14c of $d = 111^{+19}_{-18}$ pc.  This result is consistent with the close distance estimates of the Ophiuchus dark clouds ($\sim 120$~pc) determined by \citet{knude98} and \citet{degeus89}.  Additionally, \citet{loinard08} find a similar mean distance of $120 \pm 4$~pc to the Ophiuchus core using the Very Long Baseline Array (VLBA) to measure the parallaxes of radio-emitting PMS stars.  The distance of $\sim$~116 pc determined from the formal orbital parameters (Table~\ref{tab.orbpar}) is slightly closer to the VLBA value than our statistical estimate, although both are in agreement within the 1-$\sigma$ uncertainties.  Ultimately, accurate distances that map the spatial extent of the Ophiuchus clouds will be necessary for determining the luminosities, and hence ages, of young stars in the region.

\subsection{Absolute Magnitudes and Effective Temperatures}
\label{sect.MKTeff}

In addition to the distance estimate, the interferometric measurements yield the flux ratio of the components in the K-band.  Because we have insufficient information to determine the luminosities, in \S~\ref{sect.tracks} we compare our results with the theoretical PMS evolutionary tracks using H-R diagrams plotting the absolute magnitude at K, $M_K$, versus the effective temperature, $T_{\rm eff}$. The primary and secondary absolute magnitudes are given by
\begin{equation}
M_{K1\rm~or~2} = m_{K1\rm~or~2} - 5\log{d} + 5 - A_{K} 
\label{eq.magK}
\end{equation}
where $d$ is the distance in pc, $111^{+19}_{-18}$~pc, and A$_K$ the extinction at K.
To calculate the extinction at V we use 
\begin{equation}
A_V = 13.83(J-H)_{\rm observed}  -8.29(H-K)_{\rm observed} -7.43
\label{eq.AK}
\end{equation}
derived by \citet{prato03} from the reddening relations used by \citet{meyer97}.  Then, with $A_{\lambda}= (0.60/\lambda)^{1.75} A_V$ for $0.9 < \lambda < 6 \mu$m  \citep{tokunaga00}, $A_K = 0.10 A_V$.

The apparent magnitudes of the Haro 1-14c binary available in the 2MASS Catalog at J, H, and K are $8.86\pm 0.03, 8.00\pm 0.05,$ and $7.78\pm 0.03$ mag, respectively. With the measured binary flux ratio at K (Table 4), the apparent magnitudes of the primary and secondary are $m_{K1}=8.04 \pm 0.04$ and $m_{K2} = 9.47 \pm 0.14$ mag. Using the 2MASS values (which apply to the light of both components) in equation~\ref{eq.AK}, we obtain a preliminary estimate of $A_V \sim 2.64\pm0.96$ mag.  We used the PMS models calculated by Siess et al. (2000) to estimate the effect of the two components and derive a corrected visual extinction of $A_V \sim 2.9\pm1.0$ mag.  Using the relation $A_K = 0.10 A_V$, we find $A_K=0.29 \pm 0.10$ mag.  The absolute magnitudes of the primary and secondary are therefore $M_{K1} = 2.52\pm 0.04$ and $M_{K2} = 3.95\pm 0.14$ mag.  The uncertainties include only the errors in the measured K-band flux ratio and the 2MASS photometry but not the uncertainties in either the distance modulus or extinction because their values apply equally to both components.  We will display the effects of the uncertainties in the distance modulus and extinction on the H-R diagrams presented in \S\ref{sect.tracks}. 

Based on the observed spectral templates, the best fitting spectral types for  the primary and secondary of Haro 1-14c are K5 and M1.5, respectively.  To estimate the effective temperatures of the primary and secondary, we used the temperature scale adopted by \citet{hillenbrand04} and derive values of $4395 \pm 20$ K for the primary and $3548 \pm 245$ K for the secondary.  We assume uncertainties of $\pm$~1 spectral subclass.

\subsection{Comparisons with the Evolutionary Tracks}
\label{sect.tracks}

\begin{figure}
        \scalebox{0.42}{\includegraphics{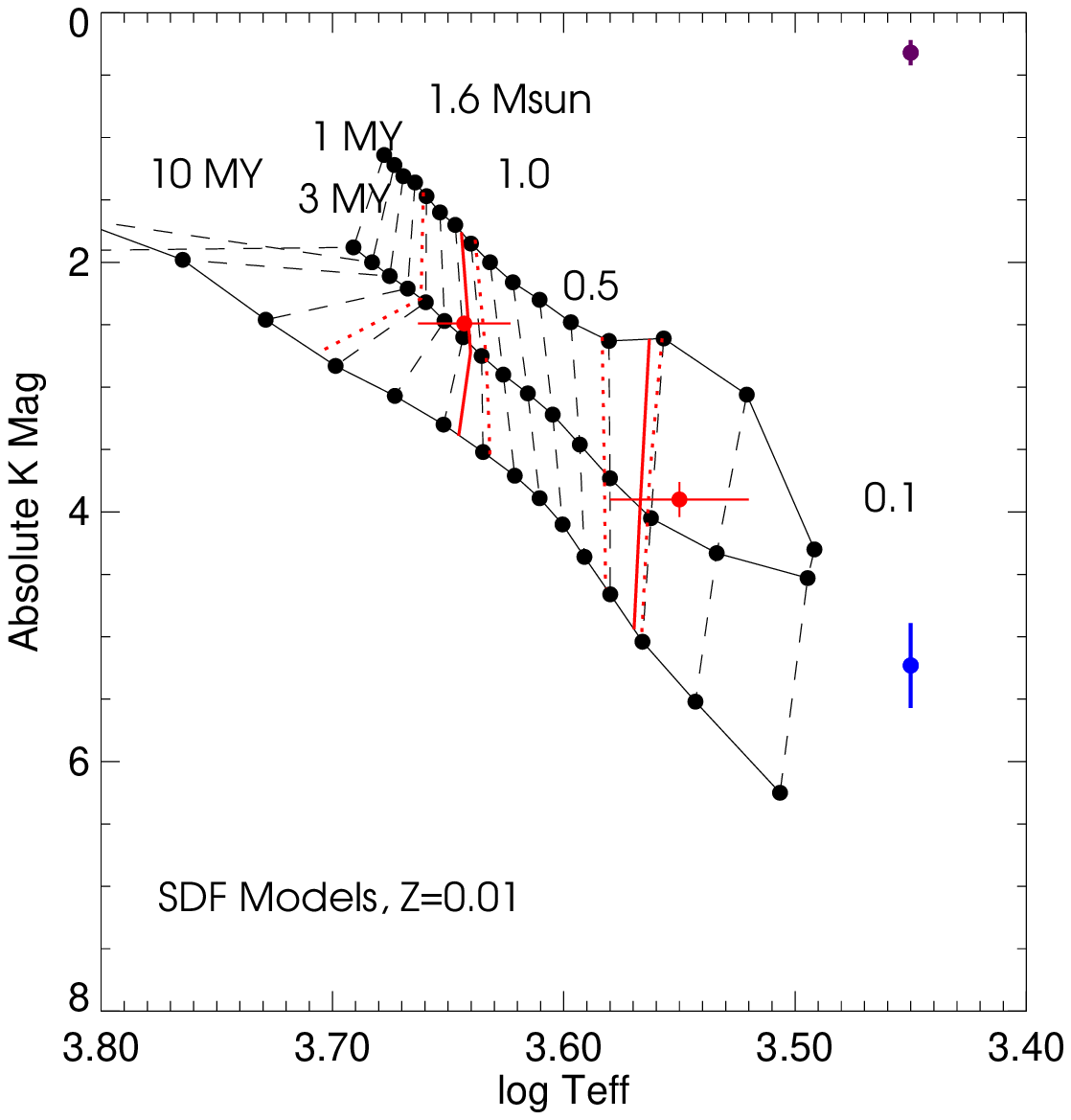}}
        \scalebox{0.42}{\includegraphics{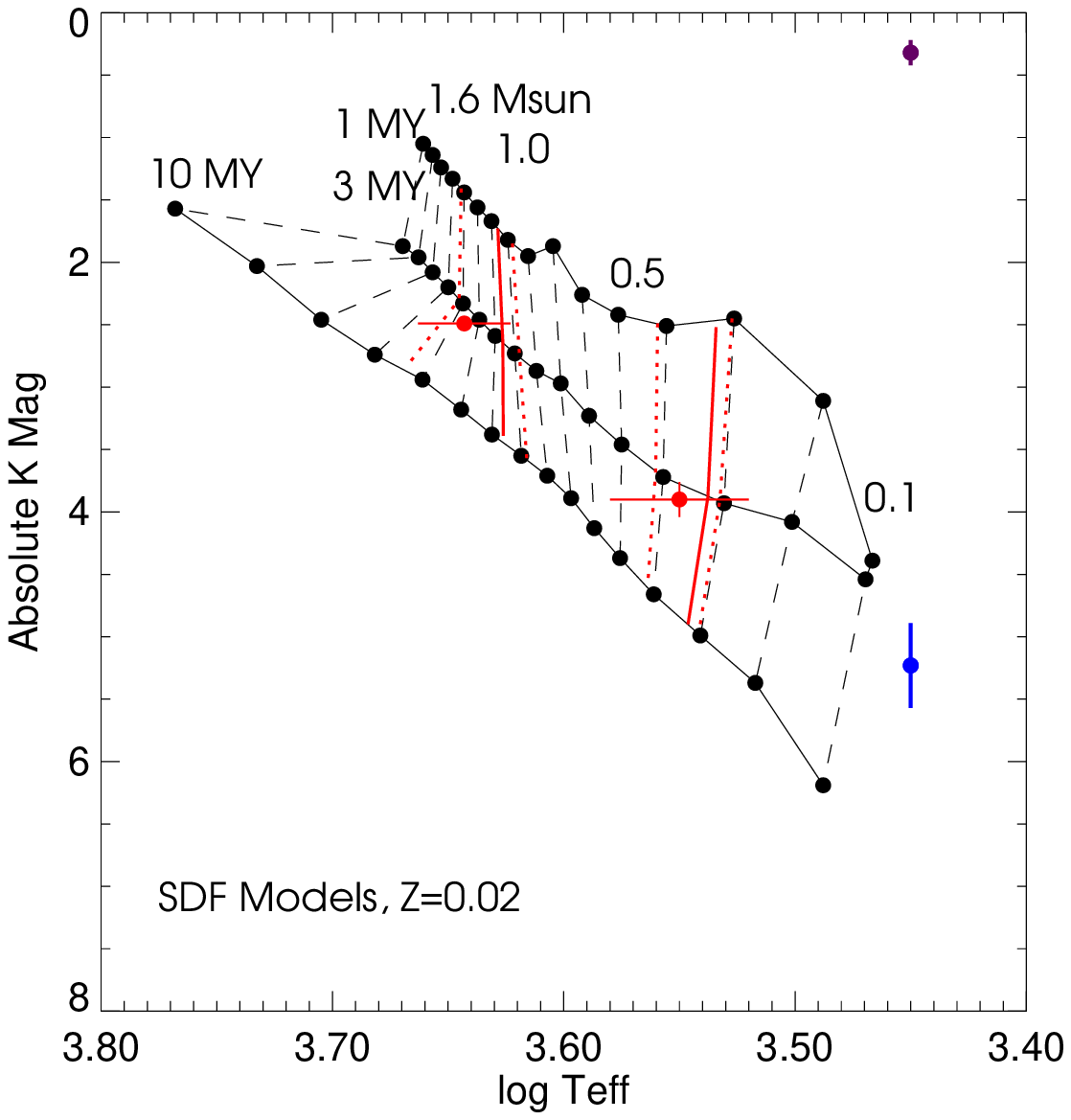}}
        \scalebox{0.42}{\includegraphics{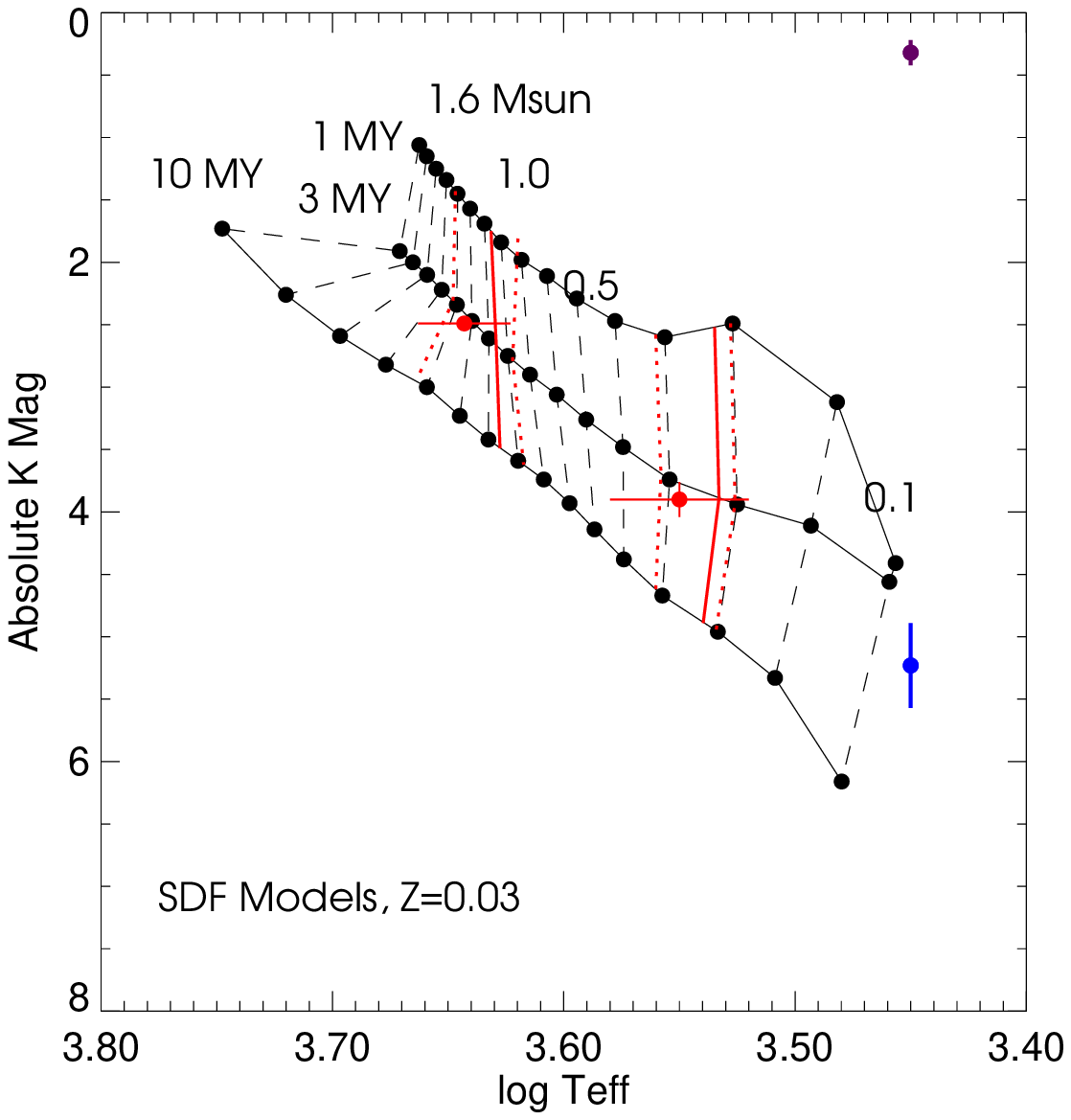}} \\
	\vspace{0.5cm}
        \scalebox{0.44}{\includegraphics{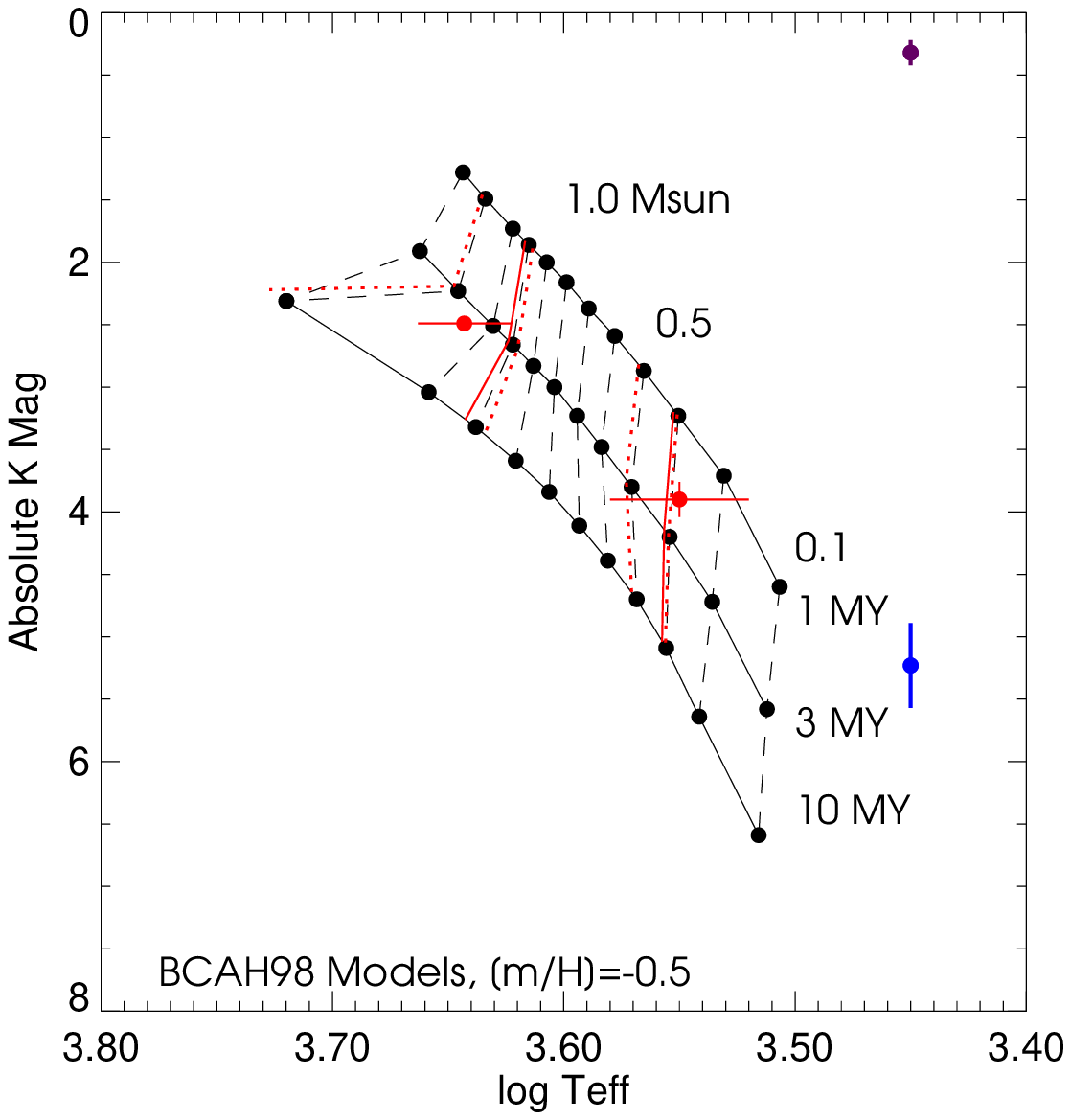}}
        \scalebox{0.44}{\includegraphics{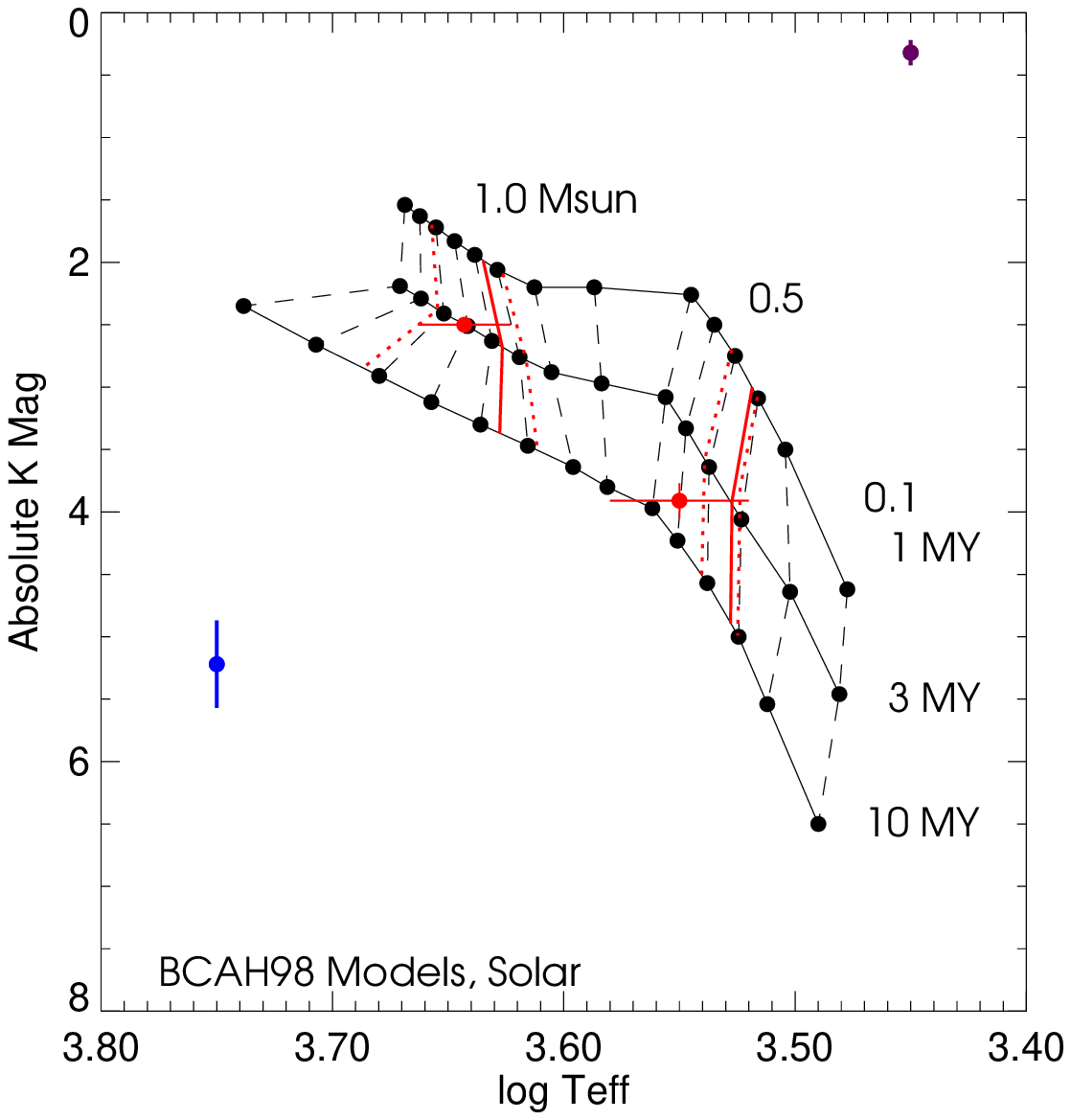}}
        \caption{{\it Top row:} Evolutionary tracks computed by SDF.  The tracks (dashed lines) cover the mass range from 0.1 to 1.6~$M_\odot$, plotted at 0.1~$M_\odot$ intervals.  The isochrones (solid lines) are plotted at ages of 1, 3, and 10 Myr.  The three panels show the SDF tracks computed at subsolar (Z=0.01; left), solar (Z=0.02; middle), and supersolar (Z=0.03; right) metallicities.  {\it Bottom row:} Evolutionary tracks computed by BCAH.  The tracks (dashed lines) cover the mass range from 0.1 to 1.4~$M_\odot$, plotted at steps of 0.1~$M_\odot$ out to 1.0~$M_\odot$ and steps of 0.2~$M_\odot$ out to 1.4~$M_\odot$.  The isochrones (solid lines) are plotted at ages of 1, 3, and 10 Myr.  The two panels show the BCAH tracks for subsolar ([m/H] = -0.5; left) and solar (right) metallicities.  In each of the panels for the BCAH and SDF tracks, the solid and dotted red lines indicate the interpolated tracks at our estimated dynamical masses of the primary and secondary of Haro 1-14c and their 1-$\sigma$ uncertainties, $M_1 = 0.96^{+0.27}_{-0.08}~M_{\odot}$ and $M_2 = 0.33^{+0.09}_{-0.02}~M_{\odot}$.  The red circles show the location of the effective temperatures and absolute K-band magnitudes.  The error bars in effective temperature represent $\pm$~1 spectral subclass and those in absolute magnitude represent only the photometric uncertainties.  The uncertainties in the distance modulus and $A_K$ are displayed by the vertical blue and purple error bars, respectively.}
\label{fig.tracks}
\end{figure}

The dynamical masses of $M_1 = 0.96^{+0.27}_{-0.08}~M_{\odot}$ and $M_2 = 0.33^{+0.09}_{-0.02}~M_{\odot}$ are in the range where precise values will make a significant contribution to testing the theoretical tracks of pre-main sequence stellar evolution.  This is particularly the case for the low mass secondary.  Moreover, the large difference in the mass of the primary and secondary will facilitate applying the coevality test to the tracks and estimating the age of the system.  Based on our statistical analysis of possible orbits, the precision of the present values of the  primary and secondary masses is about 28\% and the precision in the distance is 17\%.  In the following sections we compare our mass estimates with the evolutionary tracks of Baraffe et al. (1998, 2002; BCAH) and Siess et al. (2000; SDF).  The evolutionary tracks are plotted in Figure~\ref{fig.tracks}, where we highlight the tracks interpolated to our dynamical mass estimates of the primary and secondary of Haro 1-14c.  We overplotted the location of the effective temperatures and absolute K-band magnitudes determined for the primary and secondary (\S\ref{sect.MKTeff}).  We discuss the comparisons between each set of models in more detail below.

\subsubsection{SDF Models}

The SDF tracks are available at subsolar, solar, and supersolar metallicities (Z=0.1, 0.2, 0.3, respectively).  We retrieved the absolute K-band magnitudes from the SDF database that were computed using the temperature scale of \citet{kenyon95}.  The tracks suggest that the Haro 1-14c binary is 3-4 Myr old.  Arguably, our measurements fit the tracks for each of the three abundances within the uncertainties, but with a slight preference for the solar or supersolar tracks.

\subsubsection{BCAH Models}

The BCAH tracks are available for solar and subsolar ([m/H]=-0.5) metallicities.  For masses below 0.7~$M_\odot$, we used the tracks calculated with a mixing length parameter of 1.0; for greater masses we used the tracks with a mixing length of 1.9.  The solar and subsolar tracks indicate an age for the binary of $\sim$~3~Myr, or slightly older.  Agreement with the mass and coevality appears to be slightly better on the subsolar tracks.  

\subsubsection{Summary of Comparisons}

Across the range of metallicities considered, both the tracks of BCAH and SDF indicate an age of 3-4 Myr for the Haro 1-14c binary.  Sliding the components up and down according to the distance and extinction uncertainties in the H-R diagrams would change the binary age but not the qualitative agreement with the tracks.  Overall, the agreement between the evolutionary tracks and our measurements seems to be better for the models of SDF than for BCAH, especially when considering the coevality criterion.  

\begin{figure}
        \scalebox{1.0}{\includegraphics{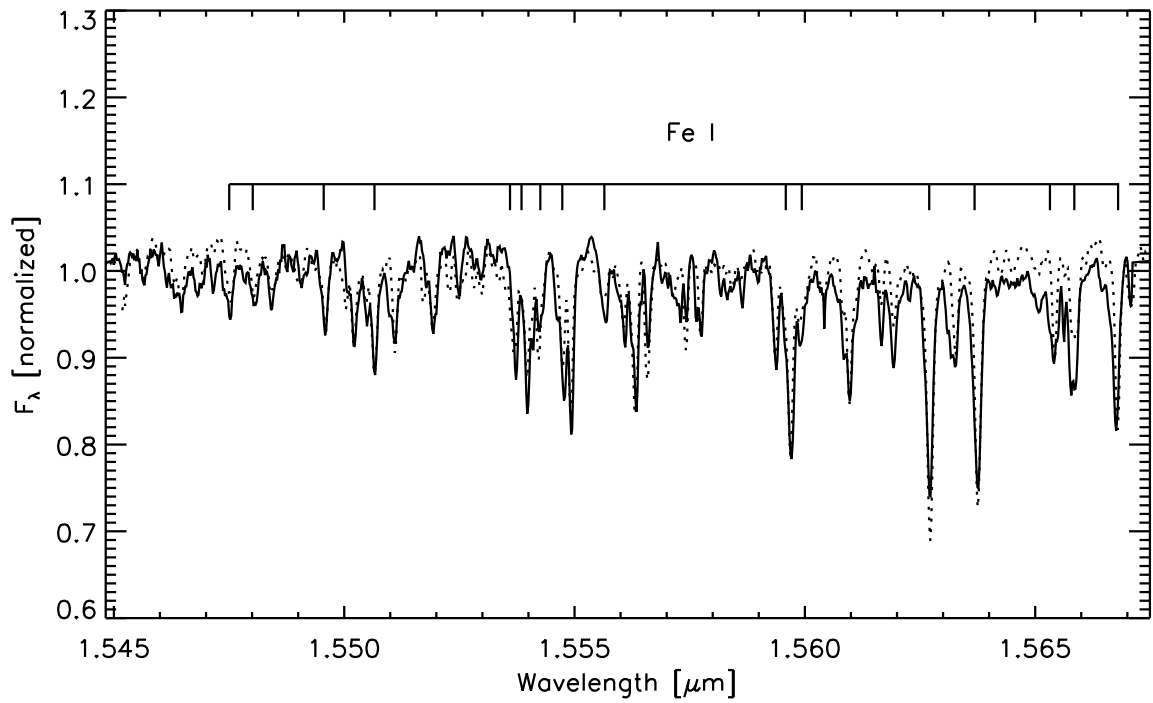}}
	\caption{NIRSPEC spectrum in order 49 of Haro 1-14c (solid line) compared with the spectral template HR 8085 (dotted line).  The location of the Fe I lines are marked in the plot.}
\label{fig.spec}
\end{figure}

The evolutionary tracks of SDF and BCAH allow for the possibility of solar, supersolar, and subsolar metallicities for Haro 1-14c.  However, based on a comparison with the solar metallicity spectral template, HR 8085, we argue that a subsolar or supersolar metallicity for Haro 1-14c is unlikely given the strength of the Fe lines across order 49 of the NIRSPEC spectrum (see Figure~\ref{fig.spec}).  Additionally, \citet{santos08} find a metallicity that is indistinguishable from solar for the Ophiuchus star-forming region.

Our age estimate of 3-4 Myr for Haro 1-14c is slightly older than the median age of 2.1 Myr determined by \citet{wilking05} for the $\rho$~Ophiuchus molecular cloud.  However, \citet{wilking05} use a distance of 150~pc in determining the stellar luminosities in their sample.  If a large portion of the stars in Ophiuchus are located at the closer distance estimate of 120~pc, the net effect would be to decrease the intrinsic luminosities and increase the age estimate of stars in the region.

From our preliminary analysis, it appears that a major limiting factor in making comparisons with the PMS evolutionary tracks lies in our ability to estimate the effective temperatures of the component stars.  Based on the spectral types of the best-fitting observed spectral templates, we used the temperature scale adopted by \citet{hillenbrand04} to estimate $T_{\rm eff}$ for the components of Haro 1-14c.  Temperatures of PMS stars determined in this way are not only affected by the $\pm$~1 spectral subclass uncertainties from the fit, but are also subject to differences in the adopted temperature scale.  For instance, the temperature scales used by \citet{kenyon95}, \citet{luhman03}, and \citet{hillenbrand04} vary by more than 200~K for stars of spectral type M2 and later.  Possible improvements in the $T_{\rm eff}$ measurements of PMS stars could be obtained from fitting synthetic spectral templates to high resolution near-infrared spectra \citep{doppmann03}, however, this may prove difficult for binaries where the spectra of the rotationally broadened components are blended.

\section{Conclusions}
\label{sect.summ}

1. Using the Keck Interferometer, we spatially resolved the orbit of the pre-main sequence binary, Haro 1-14c.  We also added two new spectroscopic measurements of the radial velocities of the primary and secondary.  

2. We computed a simultaneous orbit fit to the interferometer visibilities and the spectroscopic radial velocities.  Based on a statistical analysis of the orbital parameters that fit the data, we determined component masses of $M_1 = 0.96^{+0.27}_{-0.08}~M_{\odot}$ and $M_2 = 0.33^{+0.09}_{-0.02}~M_{\odot}$.  The secondary in Haro 1-14c is one of the lowest mass PMS stars with a dynamical measurement of its mass.

3. Our orbital measurements also yield a dynamical distance to Haro 1-14c of $111^{+19}_{-18}$~pc.  This result is consistent with the close distance estimates to the Ophiuchus molecular cloud.

4. Comparing our results with the evolutionary tracks of BCAH and SDF suggests an age of 3-4 Myr for Haro 1-14c.  

5. The tracks of SDF appear to be more consistent with our measurements of Haro 1-14c than those of BCAH, especially when considering the coevality of the components.

6. Additional interferometric observations of Haro 1-14c at well-chosen times during the orbit will substantially improve the estimate of the orbital inclination.  Not only will this provide precise values for the dynamical masses of the components, but it will also improve the distance measurement for the system.  An accurate distance is needed to derive the luminosities and place the stars on the H-R diagram.  The remaining issue that needs to be addressed in comparing dynamical mass measurements with theoretical evolutionary tracks lies in determining accurate effective temperatures for PMS binaries.

\acknowledgments

We thank the staff at the Keck Interferometer and the Michelson Science Center for their superb efforts in running the interferometer and obtaining these observations.  We are grateful to Rafael Millan-Gabet for guiding us through the early stages of planning the observations and to Rachel Akeson for carefully extracting the fringe measurements obtained in April 2005.  We are also very appreciative of Andy Boden and Anneila Sargent for providing the Haro 1-14c observations obtained during their run at the Keck Interferometer in April 2006.  We thank Josh Schleider for preparing the theoretical spectra computed by T.B. for the correlation analysis.  We also acknowledge Shay Zucker and Tsevi Mazeh for making the TODCOR routine available to us.  We thank our referee, Bo Reipurth, for a careful reading of our manuscript and for providing comments that led to significant improvements in the analysis.  M.S.'s contribution to the paper was made during his sabbatical visit at the American Museum of Natural History; M.S. thanks Mordecai Mac-Low and Ben Oppenheimer for making it possible and for their hospitality.  This work was supported in part by NSF grants AST 04-44017 (L.P.) and  06-07612 (M.S.), NASA NNX07A176G (M.S. and G.H.S.), and JPL contracts 1269936 and 1294678 (G.H.S.). The data were obtained at the Keck Observatory from time allocated to NASA through a partnership with Caltech and the University of California.  The Observatory was made possible by the generous financial support of the W.M. Keck Foundation.  We wish to recognize the Hawaiian community for the opportunity to conduct these observations from the summit of Mauna Kea. This research has made use of the software packages developed at the Michelson Science Center, data products from the Two Micron All Sky Survey, and catalogs accessible through the VizieR catalogue access tool, CDS, Strasbourg, France.

\clearpage
\begin{deluxetable}{llccccccl}
\tablewidth{0pt}
\tablecaption{Calibrators for Keck Interferometer Observation of Haro 1-14c}
\tablehead{\colhead{Object} & \colhead{Coordinates} & \colhead{SpT} & \colhead{V\tablenotemark{a}} & \colhead{K\tablenotemark{b}} & \colhead{$\pi$\tablenotemark{c}} & \colhead{Sep} & \colhead{Ang Diam} & \colhead{Epoch\tablenotemark{d}} \\
\colhead{} & \colhead{} & \colhead{} & \colhead{} & \colhead{} & \colhead{(mas)} & \colhead{($^{\circ}$)} & \colhead{(mas)} & \colhead{}}
\startdata
Haro 1-14c p & 16:31:04.36~~~-24:04:32.7  & K5V & 12.3 & 8.0 &      &      &             & 1,2,4,5 \\
Haro 1-14c s & 16:31:04.36~~~-24:04:32.7  & M1.5V &    & 9.5 &      &      &           & 1,2,4,5 \\
HD 143096  & 15 59 25.863 -30 19 57.046 & K0V &  9.6 & 7.4 & 19.2 &  9.1 & 0.17$\pm$0.07 & 1       \\
HD 143955  & 16 04 00.746 -19 06 15.841 & G5V &  9.2 & 7.5 & 15.3 &  8.1 & 0.14$\pm$0.04 & 3,4     \\
HD 147935  & 16 25 49.189 -27 49 09.193 & G5V &  9.2 & 7.6 & 17.5 &  2.6 & 0.14$\pm$0.03 & 1,2,4,5 \\
HD 151692  & 16 49 53.157 -24 26 48.976 & K3V &  9.6 & 6.9 & 29.1 &  4.3 & 0.25$\pm$0.12 & 1,4,5   \\
HD 152400  & 16 54 14.962 -28 06 53.799 & G6V &  9.2 & 7.7 & 12.9 &  6.5 & 0.13$\pm$0.02 & 1       \\
HD 155006  & 17 09 50.413 -21 47 31.502 & G6V &  9.2 & 6.4 & 10.1 &  9.7 & 0.31$\pm$0.11 & 5       \\
\hline
HD 147284  & 16 21 55.439 -24 59 28.690 & G3V &  8.8 & 7.3 & 13.5 &  2.3 & 0.16$\pm$0.05 & 3 \\
HD 148376  & 16 28 02.566 -13 06 10.488 & F7V &  8.7 & 7.2 &  9.8 & 11.1 & 0.15$\pm$0.02 & 2 \\
HD 153022  & 16 57 27.164 -12 14 04.254 & G8V &  8.8 & 5.8 &  4.6 & 13.7 & 0.43$\pm$0.20 & 2 \\
\enddata
\tablenotetext{a}{V-band magnitudes from the All-sky Compiled Catalog of 2.5 million stars (Kharchenko 2001)}
\tablenotetext{b}{K-band magnitudes from the 2MASS Catalog (Cutri et al. 2003)}
\tablenotetext{c}{Parallaxes from the Hipparcos and Tycho Catalog (Perryman et al. 1997)}
\tablenotetext{d}{Epoch: 1 = 2004 June 2; 2 = 2005 April 20; 3 = 2006 April 4; 4 = 2006 May 17; 5 = 2007 May 1}
\label{tab.cal}
\end{deluxetable}

\begin{deluxetable}{cccccccc}
\tablewidth{0pt}
\tablecaption{Haro 1-14c $V^2$ Measurements}
\tablehead{\colhead{MJD} & \colhead{Date} & \colhead{UT} & \colhead{$\lambda$} & \colhead{$V^2$} & \colhead{$\sigma_{V^2}$} & \colhead{$u$} & \colhead{$v$} \\
\colhead{} & \colhead{} & \colhead{} & \colhead{($\mu$m)} & \colhead{} & \colhead{} & \colhead{(m)} & \colhead{(m)}}
\startdata
53158.37862 &  2004 Jun  2 &  09:05:12 & 2.18 & 0.795 & 0.152 & 55.98 & 54.70 \\
53158.40510 &  2004 Jun  2 &  09:43:20 & 2.18 & 0.873 & 0.123 & 53.89 & 50.95 \\
53158.41680 &  2004 Jun  2 &  10:00:11 & 2.18 & 0.881 & 0.090 & 52.49 & 49.35 \\
53158.42913 &  2004 Jun  2 &  10:17:56 & 2.18 & 0.947 & 0.083 & 50.70 & 47.72 \\
53158.44104 &  2004 Jun  2 &  10:35:06 & 2.18 & 0.776 & 0.119 & 48.68 & 46.19 \\
53158.45318 &  2004 Jun  2 &  10:52:34 & 2.18 & 0.686 & 0.089 & 46.34 & 44.71 \\
53158.46304 &  2004 Jun  2 &  11:06:46 & 2.18 & 0.645 & 0.070 & 44.24 & 43.56 \\
53158.47340 &  2004 Jun  2 &  11:21:41 & 2.18 & 0.452 & 0.070 & 41.85 & 42.41 \\
53480.57093 &  2005 Apr 20 &  13:42:08 & 2.18 & 0.295 & 0.162 & 46.41 & 44.75 \\
53480.61274 &  2005 Apr 20 &  14:42:20 & 2.18 & 0.252 & 0.223 & 36.41 & 40.27 \\
53839.58124 &  2006 Apr 14 &  13:56:58 & 2.15 & 0.824 & 0.055 & 47.75 & 45.57 \\
53839.60523 &  2006 Apr 14 &  14:31:32 & 2.15 & 0.946 & 0.067 & 42.64 & 42.77 \\
53839.62265 &  2006 Apr 14 &  14:56:37 & 2.15 & 0.719 & 0.028 & 38.32 & 40.96 \\
53839.63442 &  2006 Apr 14 &  15:13:33 & 2.15 & 0.495 & 0.038 & 35.14 & 39.85 \\
53872.45111 &  2006 May 17 &  10:49:35 & 2.15 & 0.498 & 0.037 & 53.79 & 50.82 \\
53872.46430 &  2006 May 17 &  11:08:35 & 2.15 & 0.646 & 0.042 & 52.16 & 49.03 \\
54221.47782 &  2007 May  1 &  11:28:03 & 2.18 & 0.890 & 0.098 & 55.40 & 53.31 \\
54221.48940 &  2007 May  1 &  11:44:43 & 2.18 & 0.983 & 0.035 & 54.42 & 51.67 \\
54221.51506 &  2007 May  1 &  12:21:40 & 2.18 & 0.991 & 0.045 & 51.25 & 48.18 \\
54221.54398 &  2007 May  1 &  13:03:19 & 2.18 & 0.898 & 0.019 & 46.07 & 44.55 \\
54221.55555 &  2007 May  1 &  13:19:59 & 2.18 & 0.818 & 0.019 & 43.56 & 43.21 \\
54221.56694 &  2007 May  1 &  13:36:23 & 2.18 & 0.769 & 0.035 & 40.86 & 41.97 \\
\enddata
\label{tab.v2}
\end{deluxetable}

\begin{deluxetable}{lcccccc}
\tablewidth{0pt}
\tablecaption{Haro 1-14c Radial Velocity Measurements}
\tablehead{\colhead{UT Date} & \colhead{MJD} & \colhead{$v_1$} & \colhead{$\sigma_1$} & \colhead{$v_2$} & \colhead{$\sigma_2$} & \colhead{Phase} \\
\colhead{} & \colhead{} & \colhead{(km s$^{-1}$)} & \colhead{(km s$^{-1}$)} & \colhead{(km s$^{-1}$)} & \colhead{(km s$^{-1}$)} & \colhead{}}
\startdata
2001 Jun 1  & 52061.51 &   -4.34 &   0.59 &  -17.51 &   1.16 &  0.308  \\
2002 Jul 17 & 52472.28 &  -18.13 &   0.60 &   15.00 &   1.16 &  0.002  \\
2003 Feb 8  & 52678.67 &   -6.14 &   0.62 &  -17.49 &   1.07 &  0.350  \\
2003 Aug 10 & 52861.32 &   -9.68 &   0.54 &   -6.00 &   1.14 &  0.659  \\
2003 Sep 8  & 52890.20 &  -10.59 &   0.57 &   -4.27 &   0.94 &  0.707  \\
2004 Jan 26 & 53030.69 &  -21.67 &   0.61 &   23.74 &   1.08 &  0.945  \\
2007 Apr 30 & 54220.48 &  -21.17 &   0.58 &   26.86 &   1.07 &  0.954  \\
2007 Aug 9  & 54321.26 &   -3.55 &   0.56 &  -22.85 &   1.36 &  0.124  \\
\enddata
\label{tab.sb2}
\end{deluxetable}

\begin{deluxetable}{lc}
\tablewidth{0pt}
\tablecaption{Haro 1-14c Orbital Parameters from a Formal Analysis}
\tablehead{\colhead{Parameter} & \colhead{Value}}
\startdata
$P$ (days)        & 592.11    $\pm$  0.12            \\      
$T$ (MJD)         & 53655.51  $\pm$  0.86            \\      
$e$               & 0.6211    $\pm$  0.0070          \\     
$\omega (^\circ)$ & 232.4     $\pm$  1.2             \\      
$a$ (mas)         & $13.0^{+2.5}_{-0.7}$             \\     
$i (^\circ)$      & $75.9^{+9.3}_{-25.3}$            \\     
$\Omega (^\circ)$ & $66.^{+18}_{-12}$                \\      
$K_1$ (km/s)      & 8.73      $\pm$  0.13            \\      
$K_2$ (km/s)      & 25.53     $\pm$  0.56            \\      
$\gamma$ (km/s)   & -8.788    $\pm$  0.076           \\      
$f$(K)            & $0.268^{+0.030}_{-0.048}$        \\     
$M_1 (M_\odot)$   & $0.975^{+0.957}_{-0.073}$        \\      
$M_2 (M_\odot)$   & $0.333^{+0.327}_{-0.025}$        \\  
$d$ (pc)          & $116.^{+38}_{-21}$               \\   
$\chi^2$          & 101.86                           \\
$\chi^2_\nu$      & 1.227                            \\
\enddata
\label{tab.orbpar}
\end{deluxetable}

\begin{deluxetable}{lc}
\tablewidth{0pt}
\tablecaption{Best-fit Masses and Distance from a Statistical Analysis of Orbital Solutions}
\tablehead{\colhead{Parameter} & \colhead{Value}}
\startdata
$M_1 (M_\odot)$   &  $0.961^{+0.265}_{-0.083}$     \\      
$M_2 (M_\odot)$   &  $0.326^{+0.092}_{-0.023}$     \\  
$d$ (pc)          &  $111.^{+19}_{-18}$            \\    
\enddata
\label{tab.mass}
\end{deluxetable}

\clearpage

\end{document}